\documentclass[titlepage]{article}
\usepackage[T1]{fontenc}
\usepackage[utf8]{inputenc}
\usepackage[english]{babel}

\usepackage{geometry}
\usepackage{parskip} % Elimina sangría de todos los parrafos

\usepackage{graphicx}

\usepackage{hyperref}
\hypersetup{
	colorlinks   = true,
	citecolor    = gray,
	linkcolor	 = blue    % Change cite, ref and eqref hyperlink colors!
}
\usepackage{cite}
\usepackage[small,bf,hang]{caption}

\usepackage{amsmath,amscd,amsfonts,mathtools,amssymb,mathrsfs} %To use math symbols and all of that
\usepackage{bm} % To use boldsymbols in math mode
\usepackage{slashed} % The slashed derivative of feynman dyagrams symbol

\usepackage{multicol}
\usepackage{array}
\usepackage{multirow}
\usepackage[justification=centering]{caption} % Justification for captions
\usepackage{diagbox} % To create cells with diagonal lines separating text.

\usepackage{breqn}

 \csname
@addtoreset\endcsname{equation}{section} % Format of equation numbering

\usepackage[most]{tcolorbox}
\usepackage{color}

\usepackage{bookman} % The best font ever
\usepackage{bm}

\usepackage{marvosym,wasysym}

\newtcolorbox{resumenbox}{
	title= \large \textbf{ \hspace{.43\textwidth} Resumen},
	coltitle=black,
	fonttitle=\scshape,
	colbacktitle=black!20!white,
	arc = 4pt,
	outer arc = 4pt,
	boxrule=1pt,
	colback=white,
	colframe = black,
	boxsep = 4pt,
	left = 6pt,
	right = 6pt
}

\newtcolorbox{equbox}{
	arc = 4pt,
	outer arc = 4pt,
	boxrule=1pt,
	colback=white,
	boxsep = 2pt,
	left = 3pt,
	right = 3pt,
	width=.8\textwidth,
	colframe = black}

\newtcolorbox{boxumeninterior}{colback=white,colframe=black,boxrule=1.5pt}

\newtcolorbox{boxcuentas}{colback=white,colframe=black,boxrule=1.5pt}

\newtcolorbox{boxquationinterior}{colback=white,colframe=black,boxrule=1pt}

\newenvironment{boxquation}{\vspace{.4cm}
	\begin{boxquationinterior}
	}
	{\end{boxquationinterior} \vspace{.4cm}
}

\newtcolorbox{colorcajas}[3][]
{
	colframe = #2!25,
	colback  = #2!10,
	coltitle = #2!20!black,
	title    = #3,
	#1,
}

\newtcolorbox{capbox}{colback=white,colframe=black, colbacktitle = white,
	coltitle = black,title=Resumen,fonttitle = \Large , center title}

\usepackage{titlesec} % Allows creating custom \section's\usepackage{url}
\usepackage{tocloft}
\usepackage{setspace}

\cftsetindents{part}{0em}{0em}
\cftsetindents{section}{2em}{1.5em}
\cftsetindents{subsection}{3em}{2.5em}
\cftsetindents{subsubsection}{4em}{3.5em}

\titleformat{\part}[hang]{\bfseries \Huge \setstretch{0.1}}
{}
{0pt}
{
	\newpage
	\vspace*{0cm}
	\textcolor{gray}{ \Large{Part \thepart}}
	\vspace*{.2cm}
	\newline
	\rule{\textwidth}{.5pt}
	\vspace{.5cm}
	\newline
}
{}

\titlespacing{\part}{0em}{0cm}{2cm}

\titleformat{\section}[hang]{\bfseries \setstretch{0.1}}{}{0pt}{\Large \thesection \hspace{5pt}}
\titlespacing{\section}{0em}{1cm}{1cm}

\titleformat{\subsection}[hang]{\large\sffamily\bfseries}{}{0pt}{\large \thesubsection \hspace{5pt}}

\titleformat{\subsubsection}[hang]{\normalsize\sffamily\bfseries\color{black!60!white}}{}{2pt}{\normalsize \thesubsubsection \hspace{10pt}}

\titleformat{\paragraph}[hang]{\normalsize\sffamily\bfseries}{}{0pt}{}

\makeatother

\usepackage{fancyhdr}
\renewcommand{\subsectionmark}[1]{} % print up to section in header, not subsection,etc.
\renewcommand{\headrulewidth}{0pt} % modify the rule behind the header.

\usepackage{xargs} %allow define commands with multiple optional arguments

\newcommand{\bbr}{\mathbb{R}}

\newcommand{\go}{$ \mathcal{G} $ }
\newcommand{\got}{$ \mathcal{G}' $ }
\newcommand{\Hgen}{\mathcal{H}}

\newcommand{\ga}{\mathfrak{g}}
\newcommand{\tga}{\mathfrak{g}'}
\newcommand{\dalg}{\mathfrak{d}}

\newcommandx{\biform}[2][1=$ \ $,2=$ \ $]{ \langle #1 , #2 \rangle }

\newcommand{\cor}{\Sigma}

\newcommand{\mani}{\mathcal{M}}

\newcommand{\qq}{\quad , \quad}
\newcommand{\ODtran}{\Psi}
\newcommand{\Eext}{E}
\newcommand{\cero}{\widehat }

\renewcommand{\det}[1]{\text{Det}(#1)}

\newcommand{\DDm}{\mathcal{M}}
\newcommand{\DDn}{\mathcal{N}}

\newcommand{\DDp}{\mathcal{P}}
\newcommand{\DDq}{\mathcal{Q}}

\newcommand{\DDa}{\mathcal{I}}
\newcommand{\DDb}{\mathcal{J}}
\newcommand{\DDc}{\mathcal{K}}
\newcommand{\DDd}{\mathcal{R}}
\newcommand{\DDe}{\mathcal{S}}

\newcommand{\DDFa}{\mathcal{A}}
\newcommand{\DDFb}{\mathcal{B}}
\newcommand{\DDFc}{\mathcal{C}}
\newcommand{\DDFd}{\mathcal{D}}

\newcommand{\Dm}{\mu}
\newcommand{\Dn}{\nu}
\newcommand{\Dr}{\rho}
\newcommand{\Ds}{\sigma}

\newcommand{\Da}{\iota}
\newcommand{\Db}{\kappa}

\newcommand{\DFa}{\alpha}
\newcommand{\DFb}{\beta}
\newcommand{\DFc}{\gamma}
\newcommand{\DFd}{\delta}
\newcommand{\DFe}{\epsilon}
\newcommand{\DFf}{\xi}

\newcommand{\ddm}{M}
\newcommand{\ddn}{N}

\newcommand{\ddp}{P}
\newcommand{\ddq}{Q}

\newcommand{\dda}{I}
\newcommand{\ddb}{J}
\newcommand{\ddc}{K}
\newcommand{\ddd}{R}
\newcommand{\dde}{S}

\newcommand{\ddFa}{A}
\newcommand{\ddFb}{B}

\newcommand{\dm}{m}
\newcommand{\dn}{n}
\newcommand{\doo}{o}
\newcommand{\dpp}{p}
\newcommand{\dq}{q}

\newcommand{\da}{i}
\newcommand{\db}{j}
\newcommand{\dc}{k}
\newcommand{\dd}{r}
\newcommand{\de}{s}
\newcommand{\df}{t}

\newcommand{\dFa}{a}
\newcommand{\dFb}{b}

\newcommand{\external}[1]{\mathbf{#1}}

\newcommand{\nii}{\external{m}}
\newcommand{\nj}{\external{n}}

\newcommand{\na}{\external{i}}
\newcommand{\nb}{\external{j}}

\newcommand{\nFa}{\external{a}}
\newcommand{\nFb}{\external{b}}

\newcommand{\mol}{\mathcal{O}}

\newcommand{\Ama}{\mathbb{A}}
\newcommand{\Bma}{\mathbb{B}}
\newcommand{\Cma}{\mathbb{C}}
\newcommand{\Dma}{\mathbb{D}}
\newcommand{\Mma}{\mathbb{M}}
\newcommand{\Nma}{\mathbb{N}}
\newcommand{\Oma}{\mathbb{O}}
\newcommand{\mma}{\text{M}}
\newcommand{\nma}{\text{N}}
\newcommand{\oma}{\text{O}}
\newcommand{\Xma}{\mathbb{X}}

\def\bec{\begin{center}}
	\def\ec{\end{center}}

\def\qq{\quad\quad}

\def\be{\begin{equation}}
\def\ee{\end{equation}}
\newcommand{\beq}{\begin{equation}\begin{aligned}}
\newcommand{\eeq}{\end{aligned}\end{equation}}
\def\bea{\begin{eqnarray}}
\def\eea{\end{eqnarray}}
\def\ba{\begin{array}}
	\def\ea{\end{array}}

\allowdisplaybreaks[2]
\begin{document}

\begin{titlepage}

\renewcommand{\headrulewidth}{1pt}
\pagestyle{fancy}{\rhead{\color{brown!60!black}{\Large\Coffeecup}}}
\rfoot{\thepage}

\rightline\today
\begin{center}
\vskip 1.6cm
{\Large \bf {Generalized Dualities and Higher Derivatives}}\\
 \vskip 2.0cm
{\large {Tomas Codina and Diego Marqu\'es}}
\vskip 0.5cm

{\it Instituto de Astronom\'ia y F\'isica del Espacio (CONICET-UBA)}\\ {\it Buenos Aires, Argentina} \\[1ex]

\vskip 0.5cm

{\small \verb"{tcodina, diegomarques}@iafe.uba.ar"}

\vskip 1cm
{\bf Abstract}	
\end{center}

\vskip 0.2cm

\noindent
\begin{narrower}

{\small

Generalized dualities had an intriguing incursion into Double Field Theory (DFT) in terms of local $O(d,d)$ transformations. We review this idea and use the higher derivative formulation of DFT to compute the first order corrections to generalized dualities. Our main result is a unified expression that can be easily specified to any generalized T-duality (Abelian, non-Abelian, Poisson-Lie, etc.) or deformations such as Yang-Baxter, in any of the theories captured by the bi-parametric deformation (bosonic, heterotic strings and HSZ theory), in any supergravity scheme related by field redefinitions. The prescription allows further  extensions to higher orders. As a check we recover some previously known particular examples. }

\end{narrower}

\vskip 1.5cm

\end{titlepage}

\newpage

\tableofcontents
\thispagestyle{empty}

\section{Introduction}

Probing space-time with strings challenges the way we describe geometry. When the target space possess commuting isometries  string theory is invariant under Abelian T-duality, the statement that different backgrounds lead to the same underlying physics. Double Field Theory (DFT) \cite{Siegel, DFT} is a framework that accounts for such a generalized geometry by making Abelian T-duality a manifest symmetry, for reviews see \cite{Reviews}.

Interestingly the isometries need not be Abelian. An extension of Buscher's procedure \cite{Buscher} to the case of backgrounds with non-commuting isometries led to the so-called non-Abelian T-duality (NATD) \cite{Quevedo}. This duality can connect backgrounds with isometry groups of different dimensions, and so its action works in one direction but not reversely. This drawback was cured in \cite{Klimcik1} where the requirement of isometries as a guiding principle was abandoned. It was proposed that there must be some higher algebraic structure relating dual models that shows up only in special cases as an isometry group. This led to the idea of Poisson-Lie T-duality \cite{Klimcik1}-\cite{Klimcik3}, a generalization of Abelian and NATD. These generalized dualities are only symmetries of the classical string, and mostly work as a solution generating technique, for a review see \cite{ThompsonReview}.

Generalized dualities are not obviously captured by the symmetries of DFT, where the rigid $O(d,d)$ invariance only accounts for Abelian T-duality. To understand how generalized dualities fit into this framework, it is convenient to consider generalized Scherk-Schwarz (gSS) reductions \cite{GSS} in the context of Gauged DFT \cite{HetDFT}-\cite{GaugedDFT}. There, the background is captured by a generalized twist matrix $U \in O(d,d)$ plus a generalized dilaton shift $e^{-2\lambda} \in \mathbb{R}^+$, that locally depend on the coordinates of the internal space. The background then gauges the effective action through the fluxes generated by the duality twist. Interestingly, there is a degeneration in the space of twists that lead to the same flux configurations \cite{DualityOrbits}\footnote{The paradigmatic case is that of $SO(4)$ gaugings generated by $O(3,3)$ valued twists representing either an $S^3$ background with $H$-flux or a T-fold.}. In fact, more generally it is enough to demand that the fluxes fall into the same duality orbit, in which case the different backgrounds would lead to the same underlying physics\footnote{Even more generally, the gaugings can fall into {\it different} orbits, as happens for instance when the backgrounds are non-unimodular. We can still can make sense of the duality as connecting solutions to {\it deformed} theories, such as generalized supergravities \cite{GSE}.}. This observation was originally done in \cite{DualityOrbits}, and lies at the core of many interesting discussions on how DFT connects with generalized dualities \cite{Hassler}-\cite{InvitationPL}. We can resume it as follows:

\begin{boxquation} Generalized dualities are represented through certain local $O(d,d)$ transformations and shifts of the generalized dilaton that relate different backgrounds (duality twists in Gauged DFT) whose gaugings fall into the same duality orbit.
\end{boxquation}

In this paper we exploit the technology of DFT to compute the first order higher derivative corrections to generalized dualities. Higher derivatives are incorporated into DFT through deformations of the double local Lorentz transformations \cite{MarquesNunez}-\cite{OddStory}\footnote{There are alternative formulations in which the generalized diffeomorphisms are deformed \cite{HSZ}, and also formulations in which the Lorentz deformations are accounted for through extensions of the duality group \cite{Approach}.}. Identifying the duality covariant DFT fields with those of supergravity requires the choice of a specific double Lorentz gauge and certain higher order field redefinitions. While in the Gauged DFT sub-sector of DFT the fields transform linearly under $O(d,d)$, the T-duality transformation of the supergravity fields gets deformed by the double Lorentz transformations and the redefinitions. Interestingly, throughout this procedure the $O(d,d)$ transformations need not be rigid, and so it can be applied to generalized dualities in light of the observation made above.

In this paper we find a unified expression for first order corrections to generalized dualities. It can be easily specified to any generalized T-duality (Abelian, non-Abelian, Poisson-Lie, etc.) and deformations such as Yang-Baxter, in any of the theories captured by the higher derivative corrections to DFT (bosonic or heterotic strings and HSZ theory), in any supergravity frame related by field redefinitions.

Before introducing the original results, we intend to provide a pedagogical introduction to DFT for readers of the generalized duality community, and the other way around. Section 2 is devoted to review some relevant aspects of DFT, its flux formulation, its gauged version and the way to encode higher derivatives. We discuss there how generalized dualities fit into Gauged DFT to leading order in $\alpha'$. Section 3 discusses how generalized dualities are captured by local $O(d,d) \times \mathbb{R}^+$ transformations, and present the explicit form of the elements of this group in the case of Abelian, non-Abelian and Poisson-Lie T-duality. Section 4 contains most of the original results of this paper, combining the ideas in section 2 and 3 to generate a general formula for higher derivative corrections to generalized dualities.

Along the paper the reader will find the following results:
\begin{itemize}
    \item  Although {\it local} $O(d,d)$ transformations and $\mathbb{R}^+$ shifts of the generalized dilaton are {\it not} symmetries of DFT, some specific elements of this group transform Gauged DFT into another Gauged DFT in the {\it same} duality orbit. In certain cases when the gaugings fall into {\it distinct} duality orbits, one can still make sense of the transformation as connecting background solutions to {\it deformed} DFTs.

    \item The {\it local} $O(d,d) \times \mathbb{R}^+$ transformations that relate dual backgrounds remain uncorrected with respect to higher derivatives. The elements that generate the generalized dualities can then be {\it read} from the backgrounds to lowest order, and {\it applied} to higher-order corrected backgrounds so as to obtain the corrections to the dual background. This result is extremely powerful, as it allows to perform duality transformations to backgrounds with higher derivatives, by knowing {\it only} the transformation to lowest order.  We give the explicit form of these transformations for different generalized dualities: Abelian (\ref{Oddabelian}), non-Abelian (\ref{NATDmatrixcurved}) and PL T-duality (\ref{OddRPL}), and discuss the relation between pluralities and the notion of orbits in Gauged DFT.

    \item In the context of Gauged DFT, the duality covariant generalized fields are linearly acted on by the local $O(d,d) \times \mathbb{R}^+$ transformation that defines the generalized duality. However, when it comes to translating  this  into the language of supergravity, the Lorentz gauge fixing and field redefinitions spoil the order and  simplicity of generalized dualities in Gauged DFT, inducing higher-order corrections to the transformations of the supergravity fields. In this paper we compute these corrections in full generality in (\ref{ResultadoFinal}). The result is remarkably simple, and still general enough to account for any of the two parameters $a$ and $b$ that control the higher-order deformations ($a = 0$ or $b = 0$ is the heterotic string, $a = b$ is bosonic, $a = -b$ is HSZ), for any generalized duality (defined as connecting background solutions through local $O(d,d)$ transformations and generalized dilaton shifts), for any supergravity scheme defined by its relation to the DFT scheme.
\end{itemize}

\section{A review of Double Field Theory}\label{sec_DFT}

In this section we set the conventions to be used throughout the paper, and briefly review the frame \cite{Siegel}, \cite{frameDFT} or flux \cite{exploringDFT} formulation of DFT, it's gauged version \cite{GaugedDFT} through gSS reductions and it's first order higher-derivative extension \cite{MarquesNunez}.

We begin with some conventions. $D$ is the dimension of the full space-time, $d$ is the dimension of the internal compact space, and $n = D-d$ is the dimension of the external space. Apart from the usual curved and flat type of indices, flux compactifications involve an extra type of internal indices that we call ``algebraic'' for reason that will become clear later. Table \ref{table_convention} contains the conventions for different type of indices in different dimensions.
\begin{center}
\begin{tabular}{|c||c|c|c|}
\hline
\diagbox[width=2cm, height=1.5cm, innerleftsep=0.5cm, innerrightsep=0.1cm]{ Dim }{ Space } & \textbf{Curved} & \textbf{Algebraic} & \textbf{Flat}\\ \hline \hline
 $ \pmb{2D} $ & $ \DDm,\DDn $ & $ \DDa,\DDb $ & $ \DDFa,\DDFb $\\ \hline
 $ \pmb D $ & $ \Dm, \Dn $ & $ \Da, \Db $ & $ \DFa,\DFb $\\ \hline
 $ \pmb{2d} $ & $ \ddm,\ddn $ & $ \dda,\ddb $ & $ \ddFa,\ddFb $\\ \hline
 $ \pmb d $ & $ \dm,\dn $ & $ \da,\db $ & $ \dFa,\dFb $\\ \hline
 $ \pmb n $ & $ \nii,\nj $ & $ \na ,\nb $ & $ \nFa,\nFb $\\ \hline
\end{tabular}
\captionof{table}{Index conventions.}
\label{table_convention}
\end{center}

\subsection{Flux formulation of DFT}

Double Field Theory (DFT) incorporates T-duality as a manifest symmetry, given by the continuous global $O(D,D)$ group that preserves the symmetric metric $\eta_{\DDm \DDn}$. This metric and its inverse are used to raise a lower the $2D$ curved indices $\DDm, \DDn$ on which $O(D,D)$ acts. Duality requires that in addition to the standard space-time coordinates $X^\mu$, the theory includes dual coordinates $\widetilde X_\mu$, associated with the winding excitations of closed string theory on backgrounds with non-trivial cycles. It is then defined over a doubled space with coordinates $\Xma^\DDm = (X^\Dm,\widetilde X_\Dm)$. The double space is however constrained. One option is to impose the strong constraint which states that all fields and their products must be annihilated by the double Laplacian
\begin{equation}
\partial_\DDm\,  \partial^\DDm \dots = 0 \ .
\end{equation}
This implies that locally there is always an $O(D,D)$ transformation that rotates into a frame in which the fields depend only on half of the coordinates. A particular solution is given by demanding that nothing depends on the dual coordinates $\widetilde \partial^\Dm = 0$ in which case the section coincides with the standard $D$-dimensional space-time on which supergravity is defined. Although flux compactifications of DFT permit a relaxation of this strong constraint \cite{GaugedDFT}, as we will discuss later, in this paper we will impose the strong constraint all along.

~

There is also a local $O(1,D - 1) \times O(1,D - 1)$ symmetry usually referred to as the double Lorentz symmetry. It preserves two symmetric matrices $\eta_{\DDFa \DDFb}$ and $\mathcal{H}_{\DDFa \DDFb}$ and acts on flat $2D$ indices $\DDFa, \DDFb$ which are raised and lowered by $\eta_{\DDFa \DDFb}$.

The field content of the theory simply consists of a generalized frame  $ E_\DDm{}^\DDFa$ and a generalized dilaton $d$, that depend on the double coordinates.
The generalized frame is constrained to satisfy
\begin{equation}
\begin{aligned}
\eta_{\DDm \DDn} = E_\DDm{}^\DDFa \, \eta_{\DDFa \DDFb} \, E_{\DDn}{}^{\DDFb} \ ,
\end{aligned}
\end{equation}
and permits to define the famous generalized metric as follows
\begin{equation} \label{GenMet}
\begin{aligned}
\Hgen_{\DDm \DDn} = E_\DDm{}^\DDFa \, \mathcal{H}_{\DDFa \DDFb}\, E_{\DDn}{}^{\DDFb} \ .
\end{aligned}
\end{equation}

The $ O(D, D)$ transformations
\begin{equation}
\Psi_\DDm{}^\DDp \, \eta_{\DDp \DDq} \, \Psi_{\DDn}{}^\DDq = \eta_{\DDm \DDn} \ , \ \ \ \  \Psi \in O(D,D) \ ,
\end{equation}
act linearly on the space and fields through matrix multiplication
\begin{equation}\label{abelianODD}
\Xma'{}^{\DDm} = \Xma^\DDn \Psi_\DDn{}^\DDm \ , \ \ \  E'(\Xma')_\DDm{}^\DDFa = \Psi_\DDm{}^\DDn E(\Xma)_\DDn{}^\DDFa \ , \ \ \  d'(\Xma') = d(\Xma) \ .
\end{equation}

The double Lorentz transformations
\begin{equation}
\mol_{\DDFa}{}^\DDFc \, \eta_{\DDFc \DDFd} \, \mol_{\DDFb}{}^\DDFd = \eta_{\DDFa \DDFb}  \ , \ \ \ \mol_{\DDFa}{}^\DDFc \, \mathcal{H}_{\DDFc \DDFd} \, \mol_{\DDFb}{}^\DDFd = \mathcal{H}_{\DDFa \DDFb}  \ , \ \ \  \mol \in O(1,D-1) \times O(1,D-1) \ ,
\end{equation}
act on the fields as follows
\begin{equation}
L(E)_\DDm{}^\DDFa = E_\DDn{}^\DDFb \mol_\DDFb{}^\DDFa \ , \ \ \  L(d) = d  \ .
\end{equation}
It is convenient to define a different set of double Lorentz invariants
\begin{equation}
\begin{aligned}
P^{(\pm)}{}_{\DDFa \DDFb} \equiv \frac{1}{2} \left(\eta_{\DDFa \DDFb} \pm \Hgen_{\DDFa \DDFb}\right)  \ ,
\end{aligned}
\end{equation}
which are projectors $P^{(\pm)2} = P^{(\pm)}$ and $P^{(\pm)} P^{(\mp)} = 0$ acting on the different factors of the double Lorentz product.
We define the following index notation for future reference
\begin{equation}
P^{(+)}{}_\DDFa{}^\DDFb T_\DDFb \equiv T_{\overline \DDFa} \ , \ \ \ P^{(-)}{}_\DDFa{}^\DDFb T_\DDFb \equiv T_{\underline \DDFb} \ ,
\end{equation}
and the same holds for curved indices. It is also convenient to deal with infinitesimal double Lorentz transformations $\mol_\DDFa{}^\DDFb = \delta_\DDFa{}^\DDFb + \Lambda_\DDFa{}^\DDFb$, parameterized by antisymmetric parameters $\Lambda_{\DDFa \DDFb} = \Lambda_{[\DDFa \DDFb]}$ which are diagonal with respect to the projections, namely $\Lambda_{\overline \DDFa  \underline \DDFb} = 0$. In terms of these, the Lorentz variations of the fields read
\begin{equation}
    \delta E_{\DDm}{}^{\DDFa} = E_\DDm{}^{\DDFb} \Lambda_\DDFb{}^\DDFa \  , \ \ \  \delta d = 0 \ .
\end{equation}

On top of these symmetries, DFT is invariant under generalized diffeomorphisms, which will play a minor role in this work. Finally, there is a crucial transformation consisting in a constant generalized dilaton shift, that we will call $\mathbb{R}^+$
\begin{equation}
    e^{-2d'(\Xma')} = e^{-2\alpha} e^{-2d(\Xma)}  \ , \ \ \ \ e^{-2\alpha} \in \mathbb{R}^+ \ .
\end{equation}
This is not a strict symmetry of the action, but a rescalling, and then the equations of motion turn out to be invariant under this symmetry. This will be crucial when it comes to gauging the theory.

~

DFT is defined by an action that is fixed by invariance under the symmetries discussed so far. In the frame formulation, it can be written compactly in terms of the so called generalized fluxes
\begin{equation}\label{DFTfluxes}
\begin{aligned}
F_{\DDFa \DDFb \DDFc} &\equiv 3 \Omega_{[\DDFa \DDFb \DDFc]}  \\
F_{\DDFa} &\equiv 2 D_\DDFa d + \Omega^{\DDFb}{}_{\DDFb \DDFa} \ \ \ \   {\rm where } \ \ \ \  \Omega_{\DDFa \DDFb \DDFc} \equiv D_{\DDFa} E^{\DDn}{}_{\DDFb} E_{\DDn \DDFc}  \ , \ \ \ \  D_{\DDFa} \equiv E^\DDm{}_{\DDFa} \partial_\DDm \ .
\end{aligned}
\end{equation}
The specific form of the action and the corresponding equations of motion are irrelevant in this paper, the only important thing we need to keep in mind is that they can be written in terms of the generalized fluxes and their flat derivatives (see \cite{exploringDFT} for the two-derivative action, and \cite{OddStory} for the first order corrections in terms of fluxes). Of special importance are certain projections of the generalized fluxes that happen to appear in higher derivative Lorentz transformations, and so we define them here for future reference
\begin{equation}\label{Fpm}
    F^{(+)}{}_{\DDFa \DDFb \DDFc} = F_{\underline \DDFa \overline {\DDFb \DDFc}} \ , \ \ \ \  F^{(-)}{}_{\DDFa \DDFb \DDFc} = F_{\overline \DDFa \underline {\DDFb \DDFc}} \ .
\end{equation}

~

Connecting with supergravity requires a $GL(n) \times O(d,d)$ decomposition of $O(D,D)$. Let us show how this works in the fully uncompactified scenario $n = D$. We first impose the strong constraint and pick the solution $\widetilde \partial^\Dm = 0$, so nothing depends on the dual coordinates. Next, we propose a parameterization of the generalized frame and dilaton
\begin{equation}\label{parameterization}
 E_\DDm{}^\DDFa = \frac 1 {\sqrt{2}}
\begin{pmatrix} - Q^t_{\Dm \Dn}\,  e^{(-)\Dn \DFa} &  Q_{\Dm \Dn}\,  e^{(+)}{}^\Dn{}_\DFa\\  e^{(-)\Dm \DFa} &  e^{(+)}{}^\Dm{}_{\DFa}
\end{pmatrix} \ , \ \ \  e^{-2 d} = \sqrt{-G}e^{-2\Phi} \ ,
\end{equation}
and also the invariant matrices
\begin{equation}
\eta_{\DDm \DDn} =
\begin{pmatrix} 0 & \delta_\Dm{}^{\Dn} \\ \delta^\Dm{}_\Dn & 0  \end{pmatrix}
\ , \ \ \  \eta_{\DDFa \DDFb} = \begin{pmatrix} - g{}_{\DFa \DFb}  & 0 \\ 0 & g^{\DFa \DFb} \end{pmatrix}
\ , \ \ \  \Hgen_{\DDFa \DDFb} = \begin{pmatrix} g{}_{\DFa \DFb}  & 0 \\ 0 & g^{\DFa \DFb} \end{pmatrix} \ .
\end{equation}
Here $Q_{\Dm \Dn} \equiv G_{\Dm \Dn} + B_{\Dm \Dn} $ and  $ g_{\DFa \DFb} = \text{diag}\{-1,1,\dots,1\} $ are $D$-dimensional Minkowski matrices that raise and lower flat $D$-dimensional indices. There are two vielbeins $ e^{(\pm)}{}_{\Dm}{}^{\DFa} $ each transforming under different factors of the Lorentz group. They differ by a Lorentz transformation, and so they generate the same metric
\begin{equation}
G_{\Dm \Dn} =  e^{(\pm)}{}_\Dm{}^{\DFa}\, g_{\DFa \DFb}\,  e^{(\pm)}{}_\Dn{}^{\DFb} \ .
\end{equation}
If desired, the generalized metric can then be computed from these definitions (\ref{GenMet})
\begin{equation}
\Hgen_{\DDm \DDn} =
\begin{pmatrix}
G_{\Dm \Dn} -   B_{\Dm \Dr}  G^{\Dr \Ds}  B_{\Ds \Dn} &  B_{\Dm \Dr}  G^{\Dr \Dn} \\
-  G^{\Dm \Dr}   B_{\Dr \Dn} &  G^{\Dm \Dn}
\end{pmatrix} \ .
\end{equation}

Using the parameterization of the generalized fields we can compute the components of the generalized fluxes. In particular we show here the non-vanishing components of $F^{(\pm)}$ in (\ref{Fpm})
\begin{equation}\label{parameterizationflux}
\begin{aligned}
F^{(\pm)}{}_{\DFa \DFb}{}^\DFc &= \frac{1}{\sqrt{2}} e^{(\mp)}{}^\Dn{}_\DFa \omega^{(\pm)}{}_{\Dn \DFb}{}^\DFc \ , \ \ \ \  \omega^{(\pm)}{}_{\Dm \DFa}{}^\DFb = \omega_{\Dm \DFa}{}^\DFb(e^{(\pm)}) \pm \frac{1}{2} H_{\Dm \DFa}{}^\DFb(e^{(\pm)}) \ ,
\end{aligned}
\end{equation}
where $\omega(e^{(\pm)})$ and $H(e^{(\pm)})$ are the Levi-Civita spin connection and curvature for the two-form respectively
\begin{equation}
\begin{aligned}
\omega_{\Dm \DFa}{}^\DFb(e) &\equiv e{}^\Dn{}_\DFa \nabla_{\Dm} e_\Dn{}^\DFb \ , \ \ \ \  H_{\Dm \DFa}{}^\DFb(e) \equiv 3 \partial_{[\Dm} B_{\Dn \Dr]} e{}^\Dn{}_{\DFa} e^{ \Dr \DFb} \ ,
\end{aligned}
\end{equation}
but evaluated in $e^{(\pm)}$ instead.

~

Making contact with supergravity requires a gauge fixing. This is achieved by choosing a double Lorentz gauge in which
\begin{equation} \label{gaugefixingDFT}
e^{(+)}{}_\Dm{}^{\DFa} = e^{(-)}{}_\Dm{}^{\DFa} \equiv e_\Dm{}^{\DFa} \ ,
\end{equation}
and then locking the vielbeins to coincide with the unique vielbein that there is in supergravity. This gauge fixing breaks the double Lorenz group down to its diagonal subgroup, and on the other hand it breaks the $O(D,D)$ covariance of the generalized frame, so the  failure of $O(D,D)$ to preserve the form of the generalized frame after the gauge fixing will have to be compensated by a restoring double Lorentz transformation.

\subsection{Gauged DFT}

We now briefly review Gauged DFT \cite{HetDFT}-\cite{GaugedDFT}, which is obtained after performing a generalized Scherk-Schwarz (gSS) reduction \cite{GSS} of DFT.  The idea  is to keep the $O(D,D)$ structure of the theory, assuming an underlying $GL(n) \times O(d,d)$ decomposition, under which the coordinates split as $\Xma^\DDm= (X^\nii, \widetilde X_\nii,  Y^\ddm )$ and the strong constraint is imposed in the external space such that $\widetilde \partial^\nii = 0$. The gSS ansatz for the fields is read from the rigid $O(D,D) \times \mathbb{R}^+$ symmetries of the equations of motion, and separating the dependence on external $X$ and internal $Y$ coordinates
\begin{equation}\label{SSansaetz}
\begin{aligned}
E(X,Y)_{\DDm}{}^{\DDFa}&=U(Y)_{\DDm}{}^{\DDa}\cero \Eext(X)_{\DDa}{}^{\DDFa} \ , \ \ \  d(X,Y) = \cero d(X) + \lambda(Y) \ ,
\end{aligned}
\end{equation}
where the fields with a hat only depend on the external coordinates and correspond to the dynamical objects in Gauged DFT. The matrix $U(Y)$ is usually called twist matrix or duality twist, as it must be $O(D,D)$ valued. It maps indices of the parent DFT $\DDm, \DDn$ to indices of the effective Gauged DFT $\DDa, \DDb$, and must be trivial in the external directions
\begin{equation}\label{fluxconstraint}
\begin{aligned}
U{}^\DDm{}_{\DDa} \partial_\DDm \cero T (X) &= \delta{}^\DDm{}_\DDa \partial_{\DDm} \cero T (X)   \ ,
\end{aligned}
\end{equation}
so it is in fact an element of $O(d,d)$. Together with $\lambda (Y)$, they encode all the dependence on the double internal coordinates, and contain  the information of the compactification background.

To understand the physics behind the ansatz, it is instructive to see how it affects the generalized metric
\begin{equation}
\begin{aligned}
\Hgen(X,Y)_{\DDm \DDn} &= U(Y)_{\DDm}{}^{\DDa} \cero \Hgen(X)_{\DDa \DDb} U(Y)_{\DDn}{}^{\DDb} \ , \ \ \ \cero \Hgen(X)_{\DDa \DDb} = \cero \Eext(X)_{\DDa}{}^{\DDFa} \Hgen_{\DDFa \DDFb} \cero \Eext(X)_{\DDb}{}^{\DDFb} \ .
\end{aligned}
\end{equation}
The full background $\Hgen(X,Y)_{\DDm \DDn}$ is written as perturbations around the compactification background $ U(Y)_{\DDm}{}^{\DDa} \delta_{\DDa \DDb} U(Y)_{\DDn}{}^{\DDb}$, where the fluctuations are governed by $\cero \Hgen(X)_{\DDa \DDb}$ around $\delta_{\DDa \DDb}$, which contains the fields in the effective action of  Gauged DFT, and is fixed by it's equations of motion.

Under the gSS ansatz the generalized fluxes (\ref{DFTfluxes}) split as a sum of external and internal parts
\begin{equation}\label{DFTfluxessplit}
F_{\DDFa \DDFb \DDFc} = \cero F(X)_{\DDFa \DDFb \DDFc}  + \cero \Eext^\DDa{}_\DDFa \cero \Eext^\DDb{}_\DDFb \cero \Eext^\DDc{}_\DDFc F_{\DDa \DDb \DDc} \ , \ \ \ \ F_{\DDFa} = \cero F(X)_{\DDFa} + \cero \Eext_\DDFa{}^\DDa F_\DDa \ ,
\end{equation}
where all the dependence on the twists ends on the gaugings, defined by
\begin{equation}\label{gaugings}
\begin{aligned}
F_{\DDa \DDb \DDc} &\equiv 3 \Omega_{[\DDa \DDb \DDc]} \\
F_{\DDa} &\equiv 2 U{}^{\DDm}{}_\DDa \partial_\DDm \lambda + \Omega^{\DDb}{}_{\DDb \DDa} \ \ \ \  {\rm where} \ \ \ \ \Omega_{\DDa \DDb \DDc} \equiv U{}^{\DDm}{}_{\DDa} \partial_{\DDm} U{}^{\DDn}{}_{\DDb} U_{\DDn \DDc} \ .
\end{aligned}
\end{equation}
Invariance of the action, covariance of the equations of motion and closure of the gauge algebra leads to a set of consistency constraints
\begin{equation}
   \partial_{[\DDa} F_{\DDb \DDc \DDd]}  - \frac 3 4 F_{[\DDa \DDb}{}^{\DDe} F_{\DDc \DDd] \DDe} = 0 \  , \ \ \  \partial^\DDc F_{\DDc \DDa \DDb} + 2 \partial_{[\DDa} F_{\DDb]} - F^{\DDc} F_{\DDc \DDa \DDb} = 0 \ ,
\end{equation}
where we are defining $\partial_{\DDa} = U^{\cal M}{}_{\DDa} \partial_{\cal M}$. Interestingly, the strong constraint implies these equations, but the reserse it not true and so this is a relaxed version of the strong constraint in the internal space, which can be truly double as long as these quadratic constraints are satisfied \cite{GaugedDFT}. Normally, the gaugings $F_{\DDa}$ receive extra contributions through the gauging of a warp factor re-scaling of the Kaluza-Klein fields that arise under a $GL(n)\times O(d,d)$ decomposition. We are not assuming such a decomposition and so we will ignore this here, for a general discussion on this point we refer to \cite{GSS} and \cite{FIgaugings}. We finally point out that normally the fluxes are taken to be constant, in which case the action reduces to a lower dimensional gauged supergravity. Here we will not always assume this, as non-constant deformations are relevant when it comes to discuss certain backgrounds that arise in the context of generalized dualities.

Since the twist matrix has to be trivial in the external sector (\ref{fluxconstraint}) it can be parameterized as
\begin{equation}\label{twistU}
U_\DDm{}^\DDa =
\begin{pmatrix}
\delta_\nii{}^\na & 0 & 0 & 0\\
0 & U_\dm{}^\da & 0 & U_{\dm \da}\\
0 & 0 & \delta^\nii{}_\na & 0\\
0 & U^{\dm \da} & 0 & U^\dm{}_\da
\end{pmatrix}
\ ,  \ \ \ \ U_\ddm{}^\dda = \begin{pmatrix}
U_\dm{}^\da & U_{\dm \da}\\
U^{\dm \da} & U^\dm{}_\da
\end{pmatrix} \ ,
\end{equation}
where we defined a $2d$-dimensional internal matrix $U_M{}^I$ that has to be $O(d,d)$ valued. Then, the gaugings only have internal components
\begin{equation}
\begin{aligned}
F_{\DDa \DDb \DDc} & \longrightarrow F_{\dda \ddb \ddc}=3\Omega_{[\dda \ddb \ddc]}   \\
F_{\DDa} & \longrightarrow F_{\dda}= 2 U^M{}_I \partial_M \lambda + \Omega^J{}_{J I} \ \ \ \ {\rm where } \ \ \ \ \Omega_{\dda \ddb \ddc} \equiv U^\ddm{}_{\dda} \partial_\ddm U^\ddn{}_\ddb U_{\ddn \ddc} \ ,
\end{aligned}
\end{equation}
that satisfy their own Jacobi identities
\begin{equation} \label{quadraticconstraints}
   \partial_{[\dda} F_{\ddb \ddc \ddd]}  - \frac 3 4 F_{[\dda \ddb}{}^{\dde} F_{\ddc \ddd] \dde} = 0 \  , \ \ \  \partial^\ddc F_{\ddc \dda \ddb} + 2 \partial_{[\dda} F_{\ddb]} - F^{\ddc} F_{\ddc \dda \ddb} = 0 \ .
\end{equation}
In the effective action, all the information of the  background is encoded exclusively in the gaugings $ F_{\dda \ddb \ddc} $ and $F_I$.  Their explicit form will depend on the twist matrix $U$, which in full generality is given by \cite{exploringDFT}
\begin{equation}\label{Ugeneral}
U = \begin{pmatrix}
u & b u^{-t} \\
\beta u  & \left(1 + \beta b\right) u^{-t}
\end{pmatrix} =  \begin{pmatrix}
1 & 0 \\
\beta & 1
\end{pmatrix} \begin{pmatrix}
1 & b \\
0 & 1
\end{pmatrix} \begin{pmatrix}
u & 0 \\
0 & u^{-t}
\end{pmatrix}  \ , \ \ \  b = - b^t \ , \ \ \  \beta = - \beta^t \ .
\end{equation}
The so called geometric and non-geometric fluxes \cite{STW} in this context are simply particular components of the gaugings, and can be expressed in terms of these background  fields \cite{GSS}
\begin{equation}
F_{\da \db \dc} = H_{\da \db \dc} \ , \ \ \  F_{\da \db}{}^{\dc} = f_{\da \db}{}^{\dc} \ , \ \ \  F_{\da}{}^{\db \dc} = Q_{\da}{}^{\db \dc} \ , \ \ \  F^{\da \db \dc} = R^{\da \db \dc} \ .
\end{equation}
A priori there is no obstruction in the formalism to reach all possible orbits of gaugings (this was proved for $O(3,3)$ in \cite{DualityOrbits}) if the strong constraint is relaxed as in \cite{GaugedDFT}, although a proof is still missing in general. It was shown in \cite{DualityOrbits} that when the twists are strong constrained, they additionally satisfy
\begin{equation}
\partial_I F^I - \frac 1 2 F_I F^I +  \frac 1 {12} F^{I J K} F_{I J K} = 0\ ,
\end{equation}
which is the condition that the gaugings admit an embedding into maximal supergravity \cite{EmbeddingMaximal}. This is {\it not} a constraint of Gauged DFT. Only a subset of the allowed gaugings satisfy this condition, and so a relaxation of the strong constraint is {\it mandatory} in order to reach all duality orbits. We refer to \cite{DualityOrbits} for discussions on this point.

Let us discuss the idea of how generalized dualities are treated in the context of Gauged DFT. Consider a background coordinatized by $Y$ and characterized by $U(Y)$ and $\lambda(Y)$ with gaugings
\begin{equation}\label{F}
\begin{aligned}
F_{\dda \ddb \ddc}  &= 3 U(Y)^\ddm{}_{[\dda} \partial_\ddm U(Y)^\ddn{}_\ddb  U(Y)_{\ddn \ddc]} \\
F_{\dda}  &= 2 U(Y)^\ddm{}_{\dda} \partial_\ddm \lambda(Y)  - \partial_M U(Y)^M{}_I \ .
\end{aligned}
\end{equation}
Next consider a different (dual) background coordinatized by $Y'$ and characterized by $U'(Y')$ and $\lambda'(Y')$ with gaugings
\begin{equation}\label{Fprime}
\begin{aligned}
F'_{\dda \ddb \ddc}  &= 3 U'(Y'){}^\ddm{}_{[\dda} \partial'_\ddm U'(Y'){}^\ddn{}_\ddb  U'(Y')_{\ddn \ddc]} \\
F'_{\dda}  &= 2 U'(Y'){}^\ddm{}_{\dda} \partial'_\ddm \lambda'(Y')  - \partial'_M U'(Y'){}^M{}_I \ .
\end{aligned}
\end{equation}
When the gaugings fall into the same duality orbit, namely when there exists a constant element $h \in O(d,d)$ such that
\begin{equation} \label{fluxtransf}
F'_{I J K} = h_I{}^L h_J{}^G h_K{}^H F_{L G H} \ , \ \ \ \ F'_I = h_I{}^L F_L  \ ,
\end{equation}
then the equations of motion of Gauged DFT for the original background, and those of the dual background are related by field redefinitions. These in fact are simply $O(d,d)$ rotations of the fields in the effective action by the same elements $h$
\begin{equation} \label{effectiverotation}
        \widehat E'(X)_I{}^A = h_I{}^J \widehat E(X)_J{}^A \ , \ \ \ \widehat d'(X) = \widehat d(X) \ .
\end{equation}
The combined action of (\ref{fluxtransf}) and (\ref{effectiverotation}) leave the full generalized fluxes (\ref{DFTfluxessplit}) invariant
\begin{equation}
   F'_{\DDFa \DDFb \DDFc}  = F_{\DDFa \DDFb \DDFc}  \ , \ \ \ F'_{\DDFa} =F_{\DDFa} \ .
\end{equation}
Moreover, since these fluxes only depend on the external coordinates $X$, flat derivatives acting on them are also invariant under this transformation $(D_{\DDFa}F)' = D_{\DDFa}F$. As a result, the full Gauged DFT action and it's equations of motion remain invariant.
The resulting effective theory for both dual backgrounds is the same, and in this sense they are dual to each other. Moreover, if the external factors of the gSS  ansatz $\widehat E(X)$ and $\widehat d(X)$ satisfy the equations of motion of Gauged DFT, then the generalized duality maps a solution to a solution. It is then trivial from the point of view of Gauged DFT that generalized dualities act as a solution generating technique at the classical level. This is nicely discussed in \cite{NATDasODDlocal}.

The twists and their duals belong to different spaces with different set of coordinates, $Y$ for the original and $Y'$ for the dual.  We can think of going from one background to the other through a transformation\footnote{Abelian T-duality is a special case in which dual coordinates are related by these elements of $O(d,d)$, namely $Y' = \psi Y$. }
\begin{equation}\label{LocalTransf}
Y \to Y' \ , \ \ \ \partial \to \partial' \ , \ \ \ U(Y) \to U'(Y') = \psi(Y,Y') U(Y) \ , \ \ \ \lambda(Y) \to \lambda'(Y') = \lambda(Y) + \alpha(Y,Y')   \ ,
\end{equation}
consisting in specific {\it local} $O(d,d)$ rotations by the elements $\psi(Y,Y')$ and {\it local} generalized dilaton shifts by $\alpha(Y,Y')$
\begin{equation}\label{LocalElements}
    \psi(Y,Y') = U'(Y') U^{-1}(Y) \in O(d,d) \ , \ \ \ \ \alpha(Y,Y') = \lambda'(Y') - \lambda(Y)\ ,
\end{equation}
that connect backgrounds whose gaugings fall into the same duality orbit. It is in this sense that generalized dualities can be defined by promoting the global symmetries of DFT into local symmetries of Gauged DFT.

We can summarize how generalized dualities are captured by Gauged DFT as follows:
\begin{boxquation}
Although {\it local} $O(d,d)$ transformations and $\mathbb{R}^+$ shifts of the generalized dilaton are {\it not} symmetries of DFT, some specific elements of this group transform Gauged DFT into another Gauged DFT in the same duality orbit. \end{boxquation}

Now suppose the following scenario. We have a local $O(d,d)\times \mathbb{R}^+$ transformation connecting two backgrounds ($U$, $\lambda$) and ($U'$, $\lambda'$) that generate  gaugings that fall into {\it distinct} duality orbits. In this case, it might be possible to {\it deform} them (by modifying the twists) and force them to coincide. If the deformation on its own generates a consistent gauging, then the backgrounds can be interpreted as solutions to  {\it different} Gauged DFTs gauged by the deformations. We will see this effect explicitly when discussing particular examples of generalized dualities.

\begin{boxquation}
    The local $O(d,d)\times \mathbb{R}^+$ transformations that connect twists ($U$, $\lambda$) and ($U'$, $\lambda'$) that fall into {\it distinct} duality orbits, can sometimes be interpreted as a mapping between solutions of deformed theories.
\end{boxquation}

\subsection{Higher derivatives in DFT}\label{higherderivatives}

In this section we review how to incorporate higher-derivatives in DFT through corrections to the double Lorentz transformations \cite{MarquesNunez}. The infinitesimal first-order  in $\alpha'$ deformation is given by  the generalized Green-Schwarz transformation (antisymmetrization of projected indices exchanges the index but not the projection $[\overline \DDFa \underline \DDFb] = \frac 1 2 (\overline \DDFa \underline \DDFb - \overline \DDFb \underline \DDFa)$)
\begin{equation}\label{gGS}
\begin{aligned}
\delta_\Lambda E_\DDm{}^\DDFa = E_\DDm{}^\DDFb\left[\Lambda_\DDFb{}^\DDFa + \Lambda^{(1)}{}_\DDFb{}^\DDFa\right] \ , \ \ \
\Lambda^{(1)}{}_{\DDFb \DDFa} \equiv a D_{[ \underline \DDFb} \Lambda_{\underline \DDFc}{}^{\underline \DDFd} F^{(-)}{}_{\overline \DDFa] \underline \DDFd}{}^{\underline \DDFc} - b D_{[\overline \DDFb} \Lambda_{\overline \DDFc}{}^{\overline \DDFd} F^{(+)}{}_{\underline \DDFa] \overline \DDFd}{}^{\overline \DDFc} \ ,
\end{aligned}
\end{equation}
where $a$ and $b$ are both $\mol(\alpha')$ and interpolate between different string effective theories. The generalized dilaton remains a Lorentz scalar. This first-order correction implies that the component fields parameterizing the generalized fields under a $GL(n)\times O(d,d)$ decomposition cannot be the standard ones that transform covariantly under Lorentz transformations. Instead, they are related to those through first order Lorentz non-covariant field redefinitions. Then, when written in terms of the Lorentz covariant fields, the generalized frame is parameterized by higher derivative terms. For this reason, it is convenient to parameterize the generalized frame as follows (for concreteness we take the case $n = D$)
\begin{equation}\label{alphaparameterization}
E_\DDm{}^\DDFa = \frac 1 {\sqrt{2}}
\begin{pmatrix} - \bar Q^t_{\Dm \Dn}\,  \bar e^{(-)\Dn \DFa} &  \bar Q_{\Dm \Dn}\,  \bar e^{(+)}{}^\Dn{}_\DFa\\  \bar e^{(-)\Dm \DFa} &  \bar e^{(+)}{}^\Dm{}_{\DFa}
\end{pmatrix} \ , \ \ \  e^{-2 d} = \sqrt{-\bar G}e^{-2\bar \Phi} \ , \ \ \  \Lambda_\DDFa{}^\DDFb = \begin{pmatrix}
\bar \Lambda^{(-)}{}_\DFa{}^\DFb & 0\\
0 &  \bar \Lambda^{(+)}{}^\DFa{}_\DFb
\end{pmatrix} \ ,
\end{equation}
where the overline indicates that the components are duality covariant but not Lorentz covariant. In other words, the duality covariant fields $\bar \Psi$ are related to the Lorentz covariant ones $\Psi$ though first order redefinitions $\Delta \Psi$, namely $\bar \Psi = \Psi +  \Delta \Psi$. Note that $\bar \Psi$ is duality covariant but Lorentz non-covariant, and $\Psi$ is the opposite. The parameterization of the first-order deformation is
\begin{equation}\label{deltaE}
\begin{aligned}
\Lambda^{(1)}{}_{\DDFb}{}^\DDFa &= \begin{pmatrix}
0 &  e^{(-)}{}^\Dm{}_\DFb e^{(+)}{}^\Dn{}_\DFa \cor_{\Dm \Dn}\\
e^{(-)}{}_\Dn{}^\DFa e^{(+)}{}_\Dm{}^\DFb \cor^{\Dn \Dm}  & 0
\end{pmatrix}\\
\cor_{\Dm \Dn} &\equiv \frac{1}{4}\left(a \cor^{(-)}{}_{\Dm \Dn} + b \cor^{(+)}{}_{\Dn \Dm}\right) \ , \ \ \
\cor^{(\pm)}{}_{\Dm \Dn} \equiv \partial_{\Dm} \Lambda^{(\pm)}{}_{\DFa}{}^\DFb \omega^{(\pm)}{}_{\Dn \DFb}{}^\DFa \ .
\end{aligned}
\end{equation}
Note that because this deformation is already first-order, it is the same to put bars or not as the difference is of higher order. The corrected transformations of the $D$-dimensional fields are given by
\begin{equation}
\begin{aligned}
\delta_{\bar \Lambda} \bar e^{(+)} &= \bar e^{(+)} \bar \Lambda^{(+)} - \cor^t G^{-1} e^{(+)} \ , \ \ \  \delta_{\bar \Lambda} \bar e^{(-)} = \bar e^{(-)} \bar \Lambda^{(-)} - \cor G^{-1} e^{(-)} \ , \ \ \  \delta_{\bar \Lambda} \bar Q = - 2\cor \ ,
\end{aligned}
\end{equation}
where we have written everything in matrix notation.

When it comes to reduce this setup to supergravity one has to perform a double Lorentz transformation to a certain gauge in which the two vielbeins coincide. These transformations are {\it finite}, so we now discuss how to extract the finite version of the double Lorentz deformations from the infinitesimal ones considered above, following the strategy in \cite{BW2} closely. We aim at re-writing the transformations in terms of $\bar \Oma^{(\pm)} = 1 + \bar \Lambda^{(\pm)} + \dots$ where the dots represent higher orders in $\bar \Lambda^{(\pm)}$, such that $\bar \Oma^{(\pm)} g \bar \Oma^{(\pm)t} = g$. Since the lowest order is trivial, let us focus on the generalized Green-Schwarz transformation. To this end, consider the finite and infinitesimal transformation of the spin connections (which follows from $L(\bar e^{(\pm)}) = \bar e^{(\pm)} \Oma^{(\pm)}$)
\begin{equation}\label{Lomega}
\begin{aligned}
L(\omega^{(\pm)}{}_{\Dm \DFa}{}^\DFb) &= \Oma^{(\pm)-1}{}_\DFa{}^\DFc \Oma^{(\pm)}{}_\DFd{}^\DFb \omega^{(\pm)}{}_{\Dm \DFc}{}^\DFd + \Oma^{(\pm)-1}{}_\DFa{}^\DFc \partial_\Dm \Oma^{(\pm)}{}_\DFc{}^\DFb\\
\delta_\Lambda \omega^{(\pm)}{}_{\Dm \DFa}{}^\DFb &= - \Lambda^{(\pm)}{}_\DFa{}^\DFc \omega^{(\pm)}{}_{\Dm \DFc}{}^\DFb + \Lambda^{(\pm)}{}_\DFc{}^\DFb \omega^{(\pm)}{}_{\Dm \DFa}{}^\DFc + \partial_\Dm \Lambda^{(\pm)}{}_{\DFa}{}^\DFb \ .
\end{aligned}
\end{equation}
Using the above we take the following tour for the symmetric part of $\cor^{(\pm)}$ in (\ref{deltaE})
\begin{equation}
\cor^{(\pm)}{}_{(\Dm \Dn)} = \partial_{(\Dm} \Lambda^{(\pm)}{}_{\DFa}{}^\DFb \omega^{(\pm)}{}_{\Dn) \DFb}{}^\DFa = \delta_{\Lambda} \left(\frac{1}{2}\omega^{(\pm)}{}_{\Dm \DFa}{}^\DFb\omega^{(\pm)}{}_{\Dn \DFb}{}^\DFa \right) \to L\left(\frac{1}{2}\omega^{(\pm)}{}_{\Dm \DFa}{}^\DFb\omega^{(\pm)}{}_{\Dn \DFb}{}^\DFa \right) \ ,
\end{equation}
ending with
\begin{equation}
\cor^{(\pm)}{}_{(\Dm \Dn)} = \omega^{(\pm)}{}_{(\Dm \DFa}{}^\DFb \partial_{\Dn)} \Oma^{(\pm)}{}_\DFb{}^\DFc \Oma^{(\pm)-1}{}_\DFc{}^\DFa - \frac{1}{2} \partial_\Dm \Oma^{(\pm)-1}{}_\DFa{}^\DFb \partial_\Dn \Oma^{(\pm)}{}_\DFb{}^\DFa\ .
\end{equation}
What we did above is the following. We identified $\cor^{(\pm)}{}_{(\Dm \Dn)}$ with the infinitesimal failure of $\frac{1}{2} {\rm tr}(\omega^{(\pm)}\omega^{(\pm)})$ to remain invariant, and the arrow indicates that we now replace $\cor^{(\pm)}{}_{(\Dm \Dn)}$ by the failure of $\frac{1}{2} {\rm tr}(\omega^{(\pm)} \omega^{(\pm)})$ to be invariant under {\it finite} Lorentz transformations.

For the antisymmetric part of $\cor^{(\pm)}$ we proceed similarly. First we note that $\bar B_{\mu \nu}$ recieves a first order Lorentz transformation from the generalized Green-Schwarz term, given by $\delta_{\Lambda} \bar B_{\mu \nu} = -2 \cor_{[\mu \nu]}$, which implies that $H_{\Dm \Dn \Dr} = 3 \partial_{[\Dm} \bar B_{\Dn \Dr]}$ cannot be the three-form field strength as it is not Lorentz invariant
\begin{equation}
\delta_\Lambda H_{\Dm \Dn \Dr} =  - \frac{3 a}{2} \partial_{[\Dm}  \left( \partial_{\Dn} \Lambda^{(-)}{}_\DFa{}^\DFb \omega^{(-)}{}_{\Dr] \DFb}{}^\DFa\right) + \frac{3 b}{2} \partial_{[\Dm}  \left( \partial_{\Dn} \Lambda^{(+)}{}_\DFa{}^\DFb \omega^{(+)}{}_{\Dr] \DFb}{}^\DFa\right)\ .
\end{equation}
The failure coincides with the infinitesimal Lorentz transformation of two copies of Chern-Simons three forms
\begin{eqnarray}
\delta_\Lambda  \text{CS}^{(\pm)}{}_{\mu\nu\rho}  &=& -\partial_{[\Dm}  \left( \partial_{\Dn} \Lambda^{(\pm)}{}_\DFa{}^\DFb \omega^{(\pm)}{}_{\Dr] \DFb}{}^\DFa\right) \\
\text{CS}^{(\pm)}{}_{\mu\nu\rho} &\equiv& \omega^{(\pm)}{}_{[\Dm \DFa}{}^\DFb \partial_\Dn \omega^{(\pm)}{}_{\Dr] \DFb}{}^\DFa + \frac{2}{3} \omega^{(\pm)}{}_{[\Dm \DFa}{}^\DFb\omega^{(\pm)}{}_{\Dn \DFb}{}^\DFc\omega^{(\pm)}{}_{\Dr] \DFc}{}^\DFa \ ,
\end{eqnarray}
such that
\begin{equation}
- 6 \partial_{[\mu} \cor_{\nu\rho]} = 3 \partial_{[\mu}{ \delta_\Lambda \bar B_{\nu \rho]}} =  \delta_\Lambda H_{\mu \nu \rho} = \frac {3a} 2 \delta_\Lambda \text{CS}^{(-)}{}_{\mu\nu\rho} -\frac {3b} 2 \delta_\Lambda \text{CS}^{(+)}_{\mu\nu\rho}  \ .
\end{equation}
As before, we now consider  the {\it finite} Lorentz transformation of the Chern-Simons three-forms
\begin{eqnarray}
L\left(\text{CS}^{(\pm)}{}_{\Dm \Dn \Dr}\right) &=& \text{CS}^{(\pm)}{}_{\Dm \Dn \Dr} + \partial_{[\Dm}\left( \omega^{(\pm)}{}_{\Dn \DFa}{}^\DFb \partial_{\Dr]}\Oma^{(\pm)}{}_\DFb{}^\DFc \Oma^{(\pm)-1}{}_\DFc{}^\DFa\right) \\&& - \frac{1}{3}\partial_{[\Dm} \Oma^{(\pm)}{}_\DFa{}^\DFb \Oma^{(\pm)-1}{}_\DFb{}^\DFc \partial_\Dn \Oma^{(\pm)}{}_\DFc{}^\DFd \Oma^{(\pm)-1}{}_\DFd{}^\DFe \partial_{\Dr]} \Oma^{(\pm)}{}_\DFe{}{}^\DFf \Oma^{(\pm)-1}{}_\DFf{}^\DFa \ , \nonumber
\end{eqnarray}
and considering that the last term is closed and then locally exact, we readily arrive at
\begin{equation}
\begin{aligned}
\cor^{(\pm)}{}_{[\Dm \Dn]} &= -\omega^{(\pm)}{}_{[\Dm \DFa}{}^\DFb \partial_{\Dn]} \Oma^{(\pm)}{}_\DFb{}^\DFc \Oma^{(\pm)-1}{}_\DFc{}^\DFa + \cor^{(\pm) \text{WZW}}{}_{\Dm \Dn}\\
\partial_{[\Dm} \cor^{(\pm) \text{WZW}}{}_{\Dn \Dr]} &= \frac{1}{3}\partial_{[\Dm} \Oma^{(\pm)}{}_\DFa{}^\DFb \Oma^{(\pm)-1}{}_\DFb{}^\DFc \partial_\Dn \Oma^{(\pm)}{}_\DFc{}^\DFd \Oma^{(\pm)-1}{}_\DFd{}^\DFe \partial_{\Dr]} \Oma^{(\pm)}{}_\DFe{}{}^\DFf \Oma^{(\pm)-1}{}_\DFf{}^\DFa \ .
\end{aligned}
\end{equation}

In conclusion, the finite version of the generalized Green-Schwarz transformation on $D$-dimensional fields is as follows
\begin{boxquation}
\begin{equation}\label{alphaLorentztD}
\begin{aligned}
L(\bar e^{(+)}) &= \bar e^{(+)} \bar \Oma^{(+)} - \cor^t G^{-1} e^{(+)} \Oma^{(+)} \\
L(\bar e^{(-)}) &= \bar e^{(-)} \bar \Oma^{(-)} - \cor G^{-1} e^{(-)} \Oma^{(-)} \\
L(\bar G) &= \bar G - (\cor + \cor^t) \\
L(\bar B) &= \bar B - (\cor - \cor^t) \\
L(\bar Q) &= \bar Q - 2\cor \\
L(\bar \Phi) &= \bar \Phi -\frac{1}{2} G^{\Dm \Dn} \cor_{\Dm \Dn}  \ ,
\end{aligned}
\end{equation}
\end{boxquation}
where
\begin{eqnarray}\label{cor}
\cor_{\Dm \Dn} &=& \frac{1}{4}\left(a \cor^{(-)}{}_{\Dm \Dn} + b \cor^{(+)}{}_{\Dn \Dm}\right) \nonumber\\
\cor^{(\pm)}{}_{\Dm \Dn} &=& \partial_{\Dm} \Oma^{(\pm)}{}_\DFb{}^\DFc \Oma^{(\pm)-1}{}_\DFc{}^\DFa \omega^{(\pm)}{}_{\Dn \DFa}{}^\DFb  - \frac{1}{2} \partial_\Dm \Oma^{(\pm)-1}{}_\DFa{}^\DFb \partial_\Dn \Oma^{(\pm)}{}_\DFb{}^\DFa + \cor^{(\pm) \text{WZW}}{}_{\Dm \Dn}  \\
\partial_{[\Dm} \cor^{(\pm) \text{WZW}}{}_{\Dn \Dr]} &=& \frac{1}{3}\partial_{[\Dm} \Oma^{(\pm)}{}_\DFa{}^\DFb \Oma^{(\pm)-1}{}_\DFb{}^\DFc \partial_\Dn \Oma^{(\pm)}{}_\DFc{}^\DFd \Oma^{(\pm)-1}{}_\DFd{}^\DFe \partial_{\Dr]} \Oma^{(\pm)}{}_\DFe{}{}^\DFf \Oma^{(\pm)-1}{}_\DFf{}^\DFa \ .\nonumber
\end{eqnarray}
In Section \ref{sec_ODDalpha} we wil specify specific dependencies on these functions. Here we are using $\cor = \cor\left(\Oma^{(+)},\Oma^{(-)}, \omega^{(\pm)}(e^{(\pm)})\right)$, but later the arguments will change. We have also included the Lorentz transformation for the dilaton field which is obtained from $L(d)=d$ and its parameterization (\ref{alphaparameterization}). This result uses the strong constraint in the supergravity frame, but otherwise is completely general and holds for any choice of the parameters $a$ and $b$.

On a different page, let us comment here how this setup can be used to compute higher derivative corrections to generalized dualities. To address this question we must follow the approach in \cite{OddStory}, which is simply the gauged version of the $\alpha'$ deformed DFT \cite{MarquesNunez}. The idea is to perform a gSS reduction of DFT to first order in $\alpha'$, which interestingly proceeds in exactly the same way as in the two-derivative case. When the gSS ansatz (\ref{SSansaetz}) is adopted, the twists $U(Y)$ and $\lambda(Y)$ end up forming the exact same fluxes that gauge the action, equations of motion and gauge transformations  in the two derivative action. This is, nor the twists nor the gaugings receive higher-derivative corrections. However, because the Gauged DFT now contains higher derivatives, the effective generalized fields $\widehat E$ and $\widehat d$ now obey higher derivative equations of motion and then
\begin{equation} \label{ExternalCorr1}
\begin{aligned}
    E(X, Y) &= U(Y) \widehat E(X) = U(Y) \left(\widehat E^{(0)}(X)  + \widehat E^{(1)}(X) + \dots\right) \\
    d(X,Y) &= \widehat d(X) + \lambda(Y) = \widehat d^{(0)}(X) + \widehat d^{(1)}(X) + \dots + \lambda(Y) \ .
\end{aligned}
\end{equation}
Following the logic of how generalized dualities are captured by Gauged DFT, we can now perform the {\it local} $O(d,d)$ transformations and shifts of the generalized dilaton (\ref{LocalTransf}) and (\ref{LocalElements}) to transform the background into its dual
\begin{equation}\label{ExternalCorr2}
\begin{aligned}
    E'(X, Y') &= U'(Y') \widehat E(X) = U'(Y') \left(\widehat E^{(0)}(X)  + \widehat E^{(1)}(X) + \dots\right) \\
    d'(X,Y') &= \widehat d(X) + \lambda'(Y') = \widehat d^{(0)}(X) + \widehat d^{(1)}(X) + \dots + \lambda'(Y')\ .
\end{aligned}
\end{equation}
As before, when the dual background $U'(Y')$ and $\lambda'(Y')$ generates gaugings that fall into the same duality orbit than those of the original background, then it is guaranteed to be a solution of the $\alpha'$ corrected Gauged DFT. If instead the orbits are different, the dual background could be a solution of a {\it deformed} $\alpha'$ corrected Gauged DFT.

There is a remarkable consequence of the fact that {\it the twists receive no corrections} and that {\it all the corrections are captured by the external part of the gSS ansatz}:
\begin{boxquation}
The {\it local} $O(d,d) \times \mathbb{R}^+$ transformations that relate dual backgrounds remain uncorrected with respect to higher derivatives. Then, we can {\it read} the elements $\psi(Y,Y') = U'(Y') U^{-1}(Y)$ and $\alpha(Y,Y') = \lambda'(Y') - \lambda(Y)$ from the backgrounds to lowest order, and {\it apply} the transformation to higher-order corrected backgrounds so as to obtain the corrections of the dual background.
\end{boxquation}

\begin{figure} \center
\includegraphics[width=10 cm]{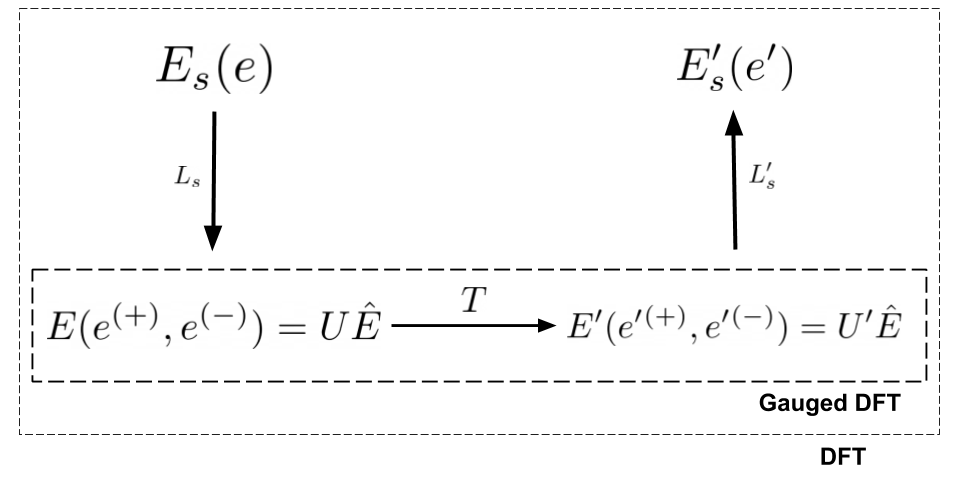}
\caption{\small{ The inner box represents Gauged DFT where the original and dual backgrounds take the gSS form. There $T$ acts linearly and receives no corrections. The double Lorentz transformations $L_s$ and $L'_s$ required to take the generalized frames to a supergravity gauge pull you out of Gauged DFT, and induce higher order corrections to generalized dualities from the perspective of supergravity.}} \label{DFTdiagram}
\end{figure}

Then, in the context of Gauged DFT the generalized frame is simply acted on linearly by the uncorrected local $T = O(d,d) \times \mathbb{R}^+$ transformation that defines the generalized duality. When it comes to make contact with supergravity,  double Lorentz transformations $L_s$ and $L'_s$ must be performed in order to take the full frames $E$ and $E'$ to a gauge in which the two vielbeins coincide (\ref{gaugefixingDFT}). This transformation is typically {\it not} allowed in Gauged DFT, and then takes you away from it. We represent the situation in Figure \ref{DFTdiagram}. If we want to explore how to go from a supergravity configuration into its dual, we must first access Gauged DFT though $L_s$ in order to take the solution into a generalized Scherk-Schwarz form, there act with $T$, and then double Lorentz transform back to the supergravity gauge in the dual picture with $L'_s$. It is through these double Lorentz transformations $L_s$ and $L'_s$ that the generalized dualities acting on supergravity backgrounds receive higher derivative corrections due to the generalized Green-Schwarz transformation.

There is a subset of double Lorentz transformations that keep you inside Gauge DFT (i.e. that preserve the gSS form of the generalized fields). These transformations are those generated by double Lorentz elements that depend only on external coordinates. These specific transformations commute with $T$ even when it is local. This is not the case of $L_s$ not $L_s'$ because these elements depend on the background which typically carries a dependence on the internal coordinates.  Figure \ref{DFTdiagram} is useful to show that starting from a corrected supergravity solution, its generalized dual is also a solution. The argument is as follows. Plugging the solution into the generalized frame  (the same story holds for the dilaton) in the supergravity gauge gives you a solution $E_s$ to the DFT equations of motion. These are covariant under generic double Lorentz transformations, and so also $E$ is a solution. Now $E$ being  a solution of DFT means that $\widehat E$  is a solution to the Gauged DFT generated by the twist $U$. The action of $T$ is to change $U$ by $U'$, but this gives you back the same Gauged DFT. So $E'$ is also a solution of DFT. Finally transforming back to the supergravity gauge with $L_s'$ (under which the DFT equations of motion are covariant) gives $E_s'$ from which the dual supergravity background can be read. So from the DFT perspective, this is a solution generating technique even at first order in $\alpha'$.

\section{$ O(D,D) $ structure of generalized dualities}\label{sec_ODDinDdimensions}

We have defined generalized dualities as the combined action of specific local $O(d,d)$ transformations and generalized dilaton shifts $\mathbb{R}^+$ that map solutions into solutions of Gauged DFTs. In this section we present the explicit form of these elements for the cases of Abelian, non-Abelian and PL T-dualities, and also Yang-Baxter deformations. In addition we discuss the embedding of these dualities into the full $O(D,D) \times \mathbb{R}^+$.

\subsection{Decompositions of $O(D,D)$}\label{glddec}

We introduce here how to decompose the group $O(D,D)$ into its subgroups $GL(D)$ (useful to deal with full $D$-dimensional solutions) and $GL(n) \times O(d,d)$ (more relevant in compactification scenarios).

\subsubsection{$GL(D)$ decomposition}

We now review the aspects of $O(D,D)$ that will be relevant to us, for more details see \cite{Porrati}. The $ O(D,D) $ group can be spanned by the matrices
\begin{equation}\label{ODDg}
\begin{aligned}
\ODtran_\bullet{}^\bullet &= \begin{pmatrix}
\Ama_\cdot{}^\cdot & \Bma_{\cdot \cdot}\\
\Cma^{\cdot \cdot} & \Dma^\cdot{}_\cdot
\end{pmatrix} \qq \Ama, \Bma, \Cma, \Dma \in \mathbb{R}^{D\times D} \qq \ODtran \eta \ODtran^t = \eta \qq \eta_{\bullet \bullet} = \begin{pmatrix}
0 &1_D\\
1_D & 0
\end{pmatrix} \ ,
\end{aligned}
\end{equation}
where the bullets represent the index structure and the $ D \times D $ matrices have to satisfy
\begin{equation}\label{ODDidentities}
\begin{aligned}
\Ama^t \Cma + \Cma^t \Ama &= \Bma^t \Dma + \Dma^t \Bma = 0 \ , \ \ \  \Ama^t \Dma + \Cma^t \Bma = 1_D\\
\Ama \Bma^t + \Bma \Ama^t &= \Cma \Dma^t + \Dma \Cma^t = 0 \ , \ \ \  \Ama \Dma^t + \Bma \Cma^t = 1_D\ .
\end{aligned}
\end{equation}
We will note the identity matrix in two different ways, depending on where the indices sit. On the one hand we have $ 1_D \equiv \delta_\Dm{}^\Dn = \delta^\Dn{}_\Dm =  \text{diag}\{1,\dots,1\}$, and on the other we will also consider the Kronecker deltas $ \delta = \delta_{\Dm \Dn}$ and $ \delta^{-1} = \delta^{\Dm \Dn}$. Identical notation will be used for dimensions other than $D$.

As it is well known, any element of the group can be decomposed as successive products of the following transformations:
\begin{itemize}
	\item \textbf{Change of basis $ A \in GL(D,\bbr) $}
	\begin{equation}\label{ATduality}
	\ODtran_{GL} = \begin{pmatrix}
	A & 0\\
	0 & A^{-t}
	\end{pmatrix} \ ,
	\end{equation}
	where $ A^{-t} \equiv (A^t)^{-1} $.
	\item  \textbf{B-shifts}
	\begin{equation}\label{BTduality}
	\ODtran_B = \begin{pmatrix}
	1_D & \Xi\\
	0 &1_D
	\end{pmatrix} \ ,
	\end{equation}
	where $ \Xi_{\Dm \Dn} = - \Xi_{\Dn \Dm}$.
	
	\item \textbf{Factorized dualities}
	
	\begin{equation}\label{TTduality}
	\ODtran_{t_\Dm} = \begin{pmatrix}
	1_D - t_\Dm & t_\Dm\\
	t_\Dm &1_D - t_\Dm
	\end{pmatrix} \ , \ \ \  (t_\Dm)_{\Dn \Dr} \equiv \delta_{\Dm \Dn}\delta_{\Dm \Dr} \ .
	\end{equation}
	
\end{itemize}

Any $ \ODtran \in O(D,D) $ can be created through succesive products of these elements. The following two transformations will be of special interest:
\begin{itemize}
	\item \textbf{Full factorized duality:}
	This transformation is obtained by applying factorized dualities over all directions
	\begin{equation}\label{fullTTduality}
	\ODtran_{f} = \begin{pmatrix}
	0 &\delta_D\\
	\delta^{-1}_D & 0
	\end{pmatrix} \ .
	\end{equation}
	
	\item \textbf{$ \beta$-shifts:}
	\begin{equation}\label{betaTduality}
	\ODtran_\beta = \begin{pmatrix}
	1_D & 0\\
	\beta &1_D
	\end{pmatrix} = \begin{pmatrix}
	0 &\delta_D\\
	\delta^{-1}_D & 0
	\end{pmatrix} \begin{pmatrix}
	1_D & \delta_D \beta \delta_D \\
	0 &1_D
	\end{pmatrix} \begin{pmatrix}
	0 &\delta_D\\
	\delta^{-1}_D & 0
	\end{pmatrix}   \ ,
	\end{equation}
	where $ \beta^{\Dm \Dn} = - \beta^{\Dn \Dm} $, and as can be seen is a product of a full factorized T-duality, a B-shift and another full factorized transformation. For this reason it is also named TsT transformation.
\end{itemize}

As explained in (\ref{abelianODD}), the $O(D,D)$ group acts linearly on the generalized frame
\begin{equation} \label{ParamFrame}
E_\DDm{}^\DDFa = \frac 1 {\sqrt{2}}
\begin{pmatrix} - Q^t_{\Dm \Dn}\,  e^{(-)\Dn \DFa} &  Q_{\Dm \Dn}\,  e^{(+)\Dn}{}_{\DFa}\\  e^{(-)\Dm \DFa} &  e^{(+)\Dm}{}_{\DFa}
\end{pmatrix} \ .
\end{equation}
We can then analyze how $ O(D,D) $ transformations act on $ D$-dimensional fields
\begin{equation}\label{ODDvielbeins}
\begin{aligned}
&T(e^{(-)}){}^{\Dm}{}_{\DFa} = \Nma^{\Dm}{}_{\Dn} e^{(-)}{}^{\Dn}{}_{\DFa} \ , \ \ \
T(e^{(+)})^{\Dm}{}_{\DFa} = \Mma^{\Dm}{}_{\Dn} e^{(+)}{}^{\Dn}{}_{\DFa}
\\ &T(Q)_{\Dm \Dn} = (\Ama Q + \Bma)_{\Dm \Dr} (\Mma^{-1})^\Dr{}_{\Dn} = (\Nma^{-t})_{\Dm}{}^\Dr (Q\Ama^t - \Bma^t)_{\Dr \Dn}\ ,
\end{aligned}
\end{equation}
where we defined
\begin{equation}
\Mma \equiv \Cma Q + \Dma \ , \ \ \  \Nma = -\Cma Q^t + \Dma \ .
\end{equation}
Using the $ O(D,D) $ identities (\ref{ODDidentities}) it can be shown that both expressions for $T(Q)$ are equivalent.

Let us now discuss how generalized T-dualities act on supergravity backgrounds to lowest order, following the route in Figure \ref{DFTdiagram}.  We first plug the supergravity background into the generalized frame in the supergravity gauge in which both vielbeins are equal (this is the starting point in the upper-left corner of Figure \ref{DFTdiagram})
\begin{equation} \label{GaugedFixedFrame}
E_s = \frac 1 {\sqrt{2}}
\begin{pmatrix} - Q^t\,  e^{-t}  \, g^{-1}&  Q\,  e^{-t}\\  e^{-t} g^{-1} &  e^{-t}
\end{pmatrix} \ .
\end{equation}
We then do a double Lorentz transformation $L_s$ to bring it to a gSS form in Gauged DFT. There, the two vielbeins are given by
\begin{equation}
 E = L_s(E_s) = E_s {\cal O}_s  = \frac 1 {\sqrt 2} \begin{pmatrix} - Q^t\,  e^{-t}  \, g^{-1}&  Q\,  e^{-t}\\  e^{-t} g^{-1} &  e^{-t}
\end{pmatrix}\begin{pmatrix} \mathbb{O}_s^{(-)} &  0 \\ 0 & \mathbb{O}_s^{(+)}
\end{pmatrix} = \frac 1 {\sqrt 2} \begin{pmatrix} - Q^t\,  e^{(-)-t}  \, g^{-1}&  Q\,  e^{(+)-t}\\  e^{(-)-t} g^{-1} &  e^{(+)-t}
\end{pmatrix}\ ,
\end{equation}
so now $e^{(\pm)}   = e\, \mathbb{O}_s^{(\pm)}$. We are now in the lower-left corner of Figure \ref{DFTdiagram}, and next we move to the right by applying the $O(D,D)$ transformation (\ref{ODDvielbeins}) $E' = T(E) = \Psi E$, which at the level of components reads
\begin{equation}
e^{(\pm)}{}' = T(e^{(\pm)}) \ , \ \ \ Q' = T(Q)\ .
\end{equation}
Finally we implement the last arrow in Figure \ref{DFTdiagram},  Lorentz transforming back with $L_s'$ to take the dual generalized frame $E'$ in Gauged DFT to the dual supergravity gauge $E_s'$
\begin{equation}\label{Eprime}
 E_s' = L_s'(E') = E' {\cal O}'_s   = \frac 1 {\sqrt 2} \begin{pmatrix} - Q'{}^t\,  e^{(-)}{}'{}^{-t}  \, g{}^{-1}&  Q'\,  e^{(+)}{}'{}^{-t}\\  e^{(-)}{}'{}^{-t} g{}^{-1} &  e^{(+)}{}'{}^{-t}
\end{pmatrix} \begin{pmatrix}\mathbb{O}_s^{(-)}{}' &  0 \\ 0 & \mathbb{O}_s^{(+)}{}'
\end{pmatrix} = \frac 1 {\sqrt{2}}
\begin{pmatrix} - Q'{}^t\,  e'{}^{-t}  \, g^{-1}&  Q'\,  e'{}^{-t}\\  e'{}^{-t} g^{-1} &  e'{}^{-t}
\end{pmatrix}\ ,
\end{equation}
where we then have $e'   = e^{(\pm)}{}'\, \mathbb{O}_s^{(\pm)}{}'$.

The composition of this sequence of transformations yields the following result for the supergravity vielbein
\begin{equation}
    e' = \mathbb{N}^{- t} e\, \mathbb{O}_s^{(-)}\mathbb{O}_s^{(-)}{}' = \mathbb{M}^{- t} e\, \mathbb{O}_s^{(+)}\mathbb{O}_s^{(+)}{}'  \  .
\end{equation}
We then see on the one hand that the Lorentz transformations are related by the fact that we are forcing the initial and dual backgrounds to be in the supergravity gauge. There is an ambiguity in how to define the supergravity gauge, because it is preserved by diagonal Lorentz transformations. We can use this freedom on both sides of the duality to set
\begin{equation}
    e^{(+)} = e \ , \ \ \ e^{(+)}{}' = e' \ , \label{diagonalLorentzgauge}
\end{equation}
so that
\begin{equation}
    \mathbb{O}_s^{(+)} = \mathbb{O}_s^{(+)}{}' = 1 \ .
\end{equation}
This choice leaves us with
\begin{equation}\label{Oorden0}
\Oma_s^{(-)} = e^{-1} e^{(-)} \ , \ \ \ \Oma_s^{(-)}{}' = e^{(-)}{}'{}^{-1} e' \ \ \Rightarrow \ \ \Oma_s^{(-)} \Oma_s^{(-)}{}' = e^{-1} \Nma^t \Mma^{-t} e = g e^{t} \Nma^{-1}  \Mma e^{-t} g^{-1}  \ ,
\end{equation}
where the last rewriting follows by using the identity
\begin{equation}\label{equationpicante}
\Mma^{-t} G \Mma^{-1} = \Nma^{-t}G\Nma^{-1} \ ,
\end{equation}
and then in this diagonal Lorentz gauge the  vielbein transforms as
\begin{equation} \label{finalvielbeintransf}
    	e' = \Mma^{-t} e \ .
\end{equation}
We could as well have chosen the gauge in which $e' = \mathbb{N}^{-t} e$, or any other one related to this by a Lorentz transformation. We will stick to the choice (\ref{diagonalLorentzgauge}) in the remainder of the paper.

From (\ref{Eprime}) we can extract $ G' $ and $ B' $ as the symmetric and anti-symmetric part of $ Q' $. Notably, using the $ O(D,D) $ identities, the transformations can be rewritten in a democratic way by defining shifted fields $ G^* $ and $ B^* $
\begin{equation}\label{Bstar}
G^* \equiv G \ , \ \ \ \ B^* \equiv B + \Dma^t \Bma + Q^t\Cma^t\Ama Q + Q^t\Cma^t\Bma - \Bma^t \Cma Q  = -B^{*t} \ .
\end{equation}
The result for the $O(D,D)$ transformations of the vielbein and two-form is:
\begin{boxquation}
	\begin{equation}\label{Los3picantes}
	\begin{aligned}
	e' &= \Mma^{-t} e  \\
	G' &= \Mma^{-t} G \Mma^{-1} = \Nma^{-t} G \Nma^{-1} \\\
	B' &= \Mma^{-t} B^* \Mma^{-1} \\
	Q' &= (\Ama Q + \Bma)\Mma^{-1} = \Mma^{-t} Q^* \Mma^{-1} \ .
	\end{aligned}
	\end{equation}
\end{boxquation}
The two ways of writing $ G' $ are equivalent due to (\ref{equationpicante}). This is not surprising because both transformations correspond to the two ways of selecting $ e' $ discussed below (\ref{finalvielbeintransf}) which are related by a Lorentz transformation, under which the metric is invariant.

The diagram in Figure \ref{DFTdiagram} applied to the dilaton field is trivial because $L(d) = d$ and $L_s'(d') = d'$.  Apart from the rigid $O(D,D)$ symmetry, the equations of motion of DFT are invariant under constant shifts of the generalized dilaton
\begin{equation}\label{gendilatonshift}
   e^{-2 d} \to e^{-2\alpha} e^{-2 d} \ , \ \ \ \ e^{-2\alpha} \in \mathbb{R}^+ \ ,
\end{equation}
and in most cases this symmetry must also be gauged for consistency. Taking into account the parameterization of the generalized dilaton (\ref{parameterization}) and that the transformation of the determinant of the metric is given by $\text{Det}(G') = \frac{\text{Det}(G)}{\text{Det}\left(\Mma\right)^2}$, we readily arrive to the $O(D,D) \times \mathbb{R}^+$ transformation of the dilaton
\begin{boxquation}
\begin{equation}\label{El4topicante}
\Phi' = \Phi - \frac{1}{2} \ln\left(\text{Det}\left(\Mma\right)\right) + \alpha \ .
\end{equation}
\end{boxquation}

Note from (\ref{Los3picantes}) and (\ref{El4topicante}) that the linear action of $ \ODtran $ over generalized tensors leads to non-linear transformations of the $ D$-dimensional fields $ G, B $ and $ \Phi $.

In the next section we will study some particular generalized dualities. From all of them, we will compute $E$ in the lower-left corner of the diagram in Figure \ref{DFTdiagram}. The most general form of the generalized vielbein  is given by (\ref{Ugeneral})
\begin{equation}\label{Egeneral}
E = U \widehat E =
\begin{pmatrix}
u & b u^{-t}\\
\beta u & (1 + \beta b)u^{-t}
\end{pmatrix}
\frac 1 {\sqrt 2}
\begin{pmatrix} - \widehat Q^t\,  \widehat e^{(-)-t}  \, g^{-1}&  Q\,  \widehat e^{(+)-t}\\  \widehat e^{(-)-t} g^{-1} &  \widehat e^{(+)-t}
\end{pmatrix} =
\frac{1}{\sqrt 2}
\begin{pmatrix} - Q^t\,  e^{(-)-t}  \, g^{-1}&  Q\,  e^{(+)-t}\\  e^{(-)-t} g^{-1} &  e^{(+)-t}
\end{pmatrix} \ ,
\end{equation}
from which one can read
\begin{equation}
\begin{aligned}
e^{(+)} &= \left[\beta u \widehat Q + (1 + \beta b)u^{-t}\right]^{-t} \widehat e^{(+)} \ , \ \ \ e^{(-)} = \left[- \beta u \widehat Q^t + (1 + \beta b)u^{-t}\right]^{-t} \widehat e^{(-)}\\ Q &= \left[u \widehat Q + b u^{-t}\right]\left[\beta u \widehat Q + (1 + \beta b)u^{-t}\right]^{-1} \ .
\end{aligned}
\end{equation}
These $e^{(\pm)}$ are exactly the ones needed to build $\Oma^{(-)}_s$ and $\Oma^{(-)}_s{}'$ (\ref{Oorden0}), considering $e = e^{(+)}$, they are given by
\begin{equation}\label{Ominusgeneral}
\begin{aligned}
\Oma^{(-)}_s &= \widehat e^{(+)-1} \left[\beta u \widehat Q + (1 + \beta b)u^{-t}\right]^{t} \left[- \beta u \widehat Q^t + (1 + \beta b)u^{-t}\right]^{-t} \widehat e^{(-)}\\
\Oma^{(-)}_s{}' &= \widehat e^{(-)-1} \left[- \beta u \widehat Q^t + (1 + \beta b)u^{-t}\right]^{t} \Nma^{t} \Mma^{-t} \left[\beta u \widehat Q + (1 + \beta b)u^{-t}\right]^{-t} \widehat e^{(+)}  \ ,
\end{aligned}
\end{equation}
where the latter can be also expressed in terms of the dual twist matrix $U'$ in terms of $u', b' $ and $\beta'$. We mentioned before that within the Gauged DFT, the allowed transformations are those that preserve the gSS form of the fields and parameters \cite{GaugedDFT},\cite{OddStory}. These transformations are inherited from the parent DFT, and act {\it only} on the external fields $\widehat E$ and $\widehat d$. Their internal coordinate dependence enters only through gaugings, and then they commute with the local elements of $T= O(D,D) \times \mathbb{R}^+$  that generate the generalized dualities. These transformations can be used to select an external double Lorentz gauge in which $\widehat e^{(+)} = \widehat e^{(-)}$.

We would like to emphasize that at no point in this section we assumed that the  $O(D,D) \times \mathbb{R}^+$ transformations are rigid, they can (and will in most cases) depend locally on the coordinates of the internal space.  We also stress that after the transformation the dual space is coordinatized by new coordinates $\Xma'$, so the effect of the transformation is not only to rotate the fields, but also to change their coordinate dependence. In the case of rigid transformations both set of coordinates are related by (\ref{abelianODD}), but in more general cases the relation is less clear. For concreteness, let us briefly discuss  the coordinate dependence of the fields
\begin{equation}\label{coordinatedependence}
E'(\mathbb{X}') = \Psi(\mathbb{X}, \mathbb{X}') E(\mathbb{X})   \ .
\end{equation}
The original background $E(\mathbb{X})$ is rotated with $\Psi(\mathbb{X}, \mathbb{X}')$ in such a way that the product $\Psi(\mathbb{X}, \mathbb{X'}) E(\mathbb{X})$ depends only on $\mathbb{X}'$. On the RHS it {\it looks like} there is some dependence on the original set of coordinates, but in reality there is not, the entire RHS is a function of $\mathbb{X}'$ only. For this reason, we are allowed replace on the RHS $\mathbb{X} \to \mathbb{X}'$ at no cost and avoid keeping track on the distinction between coordinates. Still, for clarity we will keep the distinction throughout the paper.

Let us point out that the generalized fluxes (\ref{DFTfluxes}) are only invariant $F'_{{\cal ABC}} = F[T(E)]_{{\cal ABC}} = F[E]_{{\cal ABC}}$ in the double Lorentz gauge in which the generalized fields and fluxes take the gSS form (\ref{DFTfluxessplit}), namely in Gauged DFT. Then, in that particular gauge one has the following $T$ transformations for the components (\ref{parameterizationflux})
\begin{equation}\label{Dualityomega}
\begin{aligned}
T(\omega^{(+)}) = \Nma^{-t} \omega^{(+)} \ , \ \ \ T(\omega^{(-)}) = \Mma^{-t} \omega^{(-)} \ .
\end{aligned}
\end{equation}
 This can also be obtained by direct computation from (\ref{Los3picantes}). Let us point out that these are not the dual torsionfull spin connections in the dual supergravity frame, which are given by $\omega^{(\pm)}{}' = \omega^{(\pm)}(e')$. Using $e' = e^{(-)}{}' \Oma_s^{(-)}{}' = \Nma^{-t} e \Oma_s^{(-)} \Oma_s^{(-)}{}' $, the latter are obtained as a combination of (\ref{Dualityomega}) and two Lorentz transformations (\ref{Lomega})
	\begin{equation}\label{Omegaprime}
	\begin{aligned}
	\omega^{(+)}{}' &= \Nma^{-t} \omega^{(+)} \\
	\omega^{(-)}{}' &= \Oma_s^{(-)}{}'{}^{-1} \Mma^{-t} \left(\Oma_s^{(-)-1} \omega^{(-)} \Oma_s^{(-)} + \Oma_s^{(-)-1} \partial \Oma_s^{(-)} \right) \Oma_s^{(-)}{}' + \Oma_s^{(-)}{}'{}^{-1}\partial \Oma_s^{(-)}{}' \ .
	\end{aligned}
	\end{equation}
To avoid confusion, what we are calculating are the torsionfull spin connections in the dual supergravity gauge, in terms of those in the original supergravity gauge. So, while (\ref{Dualityomega}) is a relation at the level of Gauged DFT, (\ref{Omegaprime}) is a relation at the level of supergravity.

\subsubsection{$GL(n) \times O(d,d)$ decomposition}\label{sec_reduction}

We now discuss the embedding of  $ O(d,d) $ into $O(D,D)$. To this end, the external components remain unchanged under the action of the duality group which only affects the internal space, namely
\begin{equation}\label{embedding}
\ODtran = \begin{pmatrix}
\Ama & \Bma\\
\Cma & \Dma
\end{pmatrix} \qq \Ama=\begin{pmatrix}
1_n & 0\\
0 & a
\end{pmatrix} \qq \Bma=\begin{pmatrix}
0 & 0\\
0 & b
\end{pmatrix} \qq \Cma=\begin{pmatrix}
0 & 0\\
0 & c
\end{pmatrix} \qq \Dma=\begin{pmatrix}
1_{n} & 0\\
0 & d
\end{pmatrix} \ ,
\end{equation}
with $ a, b, c, d $ being $ d\times d $ matrices. These internal matrices can be rewritten in terms of an $ O(d,d) $ object
\begin{equation}\label{psi}
\psi = \begin{pmatrix}
a & b\\
c & d
\end{pmatrix} \ ,
\end{equation}
which is the internal version of $ \ODtran $. It will be always possible to get $ \ODtran $  from $\psi$ using the trivial embedding (\ref{embedding}).

Introducing this into (\ref{Los3picantes}) and decomposing the $ D$-dimensional fields
\begin{equation}
Q_{\Dm \Dn} = \begin{pmatrix}
Q_{\nii \nj} & Q_{\nii \dn}\\
Q_{\dm \nj} & Q_{\dm \dn}
\end{pmatrix}
\qq e_{\Dm}{}^{\DFa} = \begin{pmatrix}
e_{\nii}{}^\DFa\\
e_{\dm}{}^\DFa
\end{pmatrix} \ ,
\end{equation}
we can get a component version of the transformations
\begin{boxquation}
	\begin{subequations}\label{dualitycomp}
		\begin{align}
		Q_{\nii \nj}' &=Q_{\nii \nj} - Q_{\nii \dpp} (\mma^{-1}c)^{\dpp \dq} Q_{\dq \nj} \qq &Q_{\dm \dn}'& = \left(a_\dm{}^\dpp Q_{\dpp \dq} + b_{\dm \dq}\right) (\mma^{-1})^{\dq}{}_\dn\label{dualitycompint}\\
		Q_{\nii \dn}' &= Q_{\nii \dpp} (\mma^{-1})^\dpp{}_\dn \qq
		&Q_{\dm \nj}'& = \left(a_\dm{}^\dpp - Q'_{\dm \dq}c^{\dq \dpp}\right)Q_{\dpp \nj}  \\
		e'_{\nii}{}^{\DFa} &= e_{\nii}{}^{\DFa} - Q_{\dpp \nii} c^{\dq \dpp} (\mma^{-t})_\dq{}^\doo e_{\doo}{}^{\DFa} \qq
		&e'_{\dm}{}^{\DFa}& = (\mma^{-t})_\dm{}^\dpp e_{\dpp}{}^{\DFa} \ ,
		\end{align}
	\end{subequations}
\end{boxquation}
where we defined the internal version of $ \Mma $
\begin{equation}
\mma \equiv (c Q_d + d) \ ,
\end{equation}
where $ Q_d $ is the matrix notation for the internal $d\times d$ components of $ Q $, namely $ Q_{\dm \dn} $.
To complete the picture, we notice that $ \text{Det}(\Mma) = \text{Det}(\mma)  $ and so
\begin{boxquation}
\begin{equation}\label{DualityPhi}
\Phi' = \Phi - \frac{1}{2} \ln\left(\text{Det}\left(\mma\right)\right) + \alpha \ .
\end{equation}
\end{boxquation}

The equations (\ref{dualitycomp}) and (\ref{DualityPhi}) relate different backgrounds connected by local $ O(d,d) \times \mathbb{R}^+$ transformations.

\subsection{Generalized Dualities}\label{sec_generalizeddualities}

In this section we give a brief review of generalized T-dualities and their embedding into $O(D,D)\times \mathbb{R}^+$. Before moving to a case by case study, we first introduce a common starting point to set the notation.  Let us emphasize that the examples discussed here do not exhaust the possibilities of generalized dualities that can be captured by Gauged DFT.

Consider a group $\cal G$ acting {\it freely} on a manifold $ M $. This means that given $g \in {\cal G}$ and $p \in M$, if $g \cdot p = p$ then $g =e$ is the identity element. This permits to take a set of adapted coordinates on the target space $ (X,g) $ where $ g \in {\cal G}$, and $X^\nii$ are the spectator fields (or external coordinates) that label the orbits of $\cal G$. As we explained before, and will discuss largely in this section, generalized dualities are represented by certain local $O(D,D) \times \mathbb{R}^+$ transformations that act exclusively on the twists that contain the information of the internal background. These are independent of the external coordinates, which then  play no role in identifying the $O(D,D) \times \mathbb{R}^+$ elements associated to the generalized dualities. They do however play a major role when it comes to computing higher derivative corrections (as discussed around (\ref{ExternalCorr1})-(\ref{ExternalCorr2})), but we will concentrate on that in the  next section. When the expectator fields are frozen to a trivial value, the action of $\cal G$ on $ M $ becomes {\it transitive}, meaning that any two points $ p_1, p_2 \in M $ are always connected trough some $g \in G$ such that $g \cdot p_1 = p_2$. In this case all the orbits become isomorphic to the manifold $ M $ itself and so we have a group manifold $ M = {\cal G}$, such that $ g \in  {\cal G}$ are the points in $ M $ parameterized with coordinates $ Y^\dm $.

Given the Lie algebra $ \ga $ of $\cal G$
\begin{equation}\label{structureconstants}
\left[t_\da, t_\db\right] = f_{\da \db}{}^\dc \, t_\dc \ , \ \ \  f_{[\da \db}{}^\dd f_{\dc]\dd}{}^\de = 0 \ ,
\end{equation}
the free and transitive right-action of $\cal G$ on $M$ is carried by left-invariant vector fields $ k_i \in \ga $ that transform the coordinates as
\begin{equation}\label{coordinate_change}
Y^\dm \to Y'{}^m = Y^\dm + \delta Y^\dm = Y^\dm + \epsilon^\da k_\da{}^\dm \ .
\end{equation}
The effect on the group element $g' = g + \delta g$ can be obtained in two different but equivalent ways: through the right action on $g' = g\, e^{\epsilon^\da t_\da} $ or by a change in the coordinates (\ref{coordinate_change}). This gives the relations
\begin{equation} \label{shiftsdeg}
\begin{aligned}
\delta g = g \epsilon^\da t_\da = \partial_\dm g \epsilon^\da  k_\da{}^\dm  \ .
\end{aligned}
\end{equation}
We now define the following quantities
\begin{equation}\label{adLR}
\text{Ad}_{g^{-1}} t_\da = g^{-1} t_\da g = a_\da{}^\db t_\db \ ,  \ \ \ L_\dm{}^\da = \left(g^{-1} \partial_\dm g\right)^\da \ , \ \ \  R_\dm{}^\da = \left(\partial_\dm g g^{-1}\right)^\da \ .
\end{equation}
The first is the adjoint action of $g$ defined by matrices $a_i{}^j$, and the last two are the left and right invariant one-forms, respectively. It is easy to see from (\ref{shiftsdeg}) that the following relations hold
\begin{equation} \label{akLRidentities}
 k_\da{}^\dm L_\dm{}^\db = \delta_\da{}^\db \ , \ \ \  k_i{}^\dm R_\dm{}^j = (a^{-1})_\da{}^\db \ , \ \ \  R_\dm{}^\db a_\db{}^\da = L_\dm{}^\da \ ,
\end{equation}
which in turn imply that the left and right invariant one-forms satisfy the Maurer-Cartan equations
\begin{equation}
d L^\da = - \frac{1}{2} f_{\db \dc}{}^\da L^\db \wedge L^\dc \ , \ \ \  d R^\da = \frac{1}{2} f_{\db \dc}{}^\da R^\db \wedge R^\dc \ .
\end{equation}
The duals of the left and right invariant one-forms are respectively left $ L_\da{}^\dm$ and right $ R_\da{}^\dm$ invariant vectors. From (\ref{akLRidentities}) we see that $  L_\da{}^\dm = k_\da{}^\dm $. The Maurer-Cartan equations lead to algebraic conditions on the vector fields
\begin{equation}\label{leftrightbrackets}
\begin{aligned}
\mathcal{L}_{k_\da} k_\db = [{k_\da},\, k_\db] = f_{\da \db}{}^\dc k_\dc \ , \ \ \  \mathcal{L}_{R_\da} R_\db =  - f_{\da \db}{}^\dc R_\dc \ , \ \ \  \mathcal{L}_{k_\da} R_\db = 0 \ ,
\end{aligned}
\end{equation}
where $\mathcal{L}$ is the Lie derivative and $[\, , ]$ the Lie bracket. The last identity follows from noticing that $R_\da = a_\da{}^\db k_\db $ and $\partial_\dm a_\da{}^\db = -L_\dm{}^\dc a_\da{}^\dd f_{\dc \dd}{}^\db$.

\subsubsection{Abelian T-duality}\label{AbelianT}

This is the simplest case of a generalized duality that relates backgrounds with Abelian isometries. When the original background posses $ d $ Abelian isometries, there is a set of commuting killing vectors $ k_\da $
\begin{equation} \label{Yshift}
\left[k_\da,k_\db \right] = 0 \ , \ \ \ \  Y^{\dm} \to Y'{}^\dm =  Y^{\dm} + \epsilon^\da k_\da{}^\dm \ .
\end{equation}
The Abelian algebra  $f_{i j }{}^k = 0$ allows to choose adapted coordinates for which all fields are independent of $Y^m$ and the compactified sigma model takes the form
\begin{equation}\label{reduced sigma}
S = \int \mathrm{d}^2\sigma \ \left[ \partial_+ X^\nii \partial_- X^\nj Q_{\nii \nj} + \partial_+ Y^\dm \partial_- X^\nj Q_{\dm \nj} + \partial_+ X^\nii \partial_-Y^\dn Q_{\nii \dn} + \partial_+Y^\dm \partial_-Y^\dn Q_{\dm \dn} \right]  \ ,
\end{equation}
where $ Q = G + B$ contains the different components of the metric and B-field which depend on $ X^\nii $ only. As explained before, for our purposes we could very well freeze the expectator fields $X^\nii$ and restrict attention to the internal sector, but we will keep track of them for the moment. Here we are neglecting the dilaton coupling which is going to be treated separately. The transformation (\ref{Yshift}) is a symmetry of the sigma-model as long as the fields satisfy  the isometry conditions
\begin{equation}\label{isometry}
\mathcal{L}_{k} Q = 0 \ .
\end{equation}

The way Abelian T-dualities emerge as symmetries was discussed by  Buscher \cite{Buscher}. Beginning with the Lagrangian (\ref{reduced sigma}) one follows a 3-step recipe:
\begin{description}
	\item[\textbf{(1)}] Gauge the global isometries, and then pick a gauge in which $ Y^\dm = 0 $
	\begin{equation}
	\partial_{\pm} Y^\dm \rightarrow D_{\pm} Y^\dm \equiv \partial_{\pm} Y^\dm  + A_{\pm}^\dm \rightarrow A_{\pm}^\dm  \ .
	\end{equation}
 	\item[\textbf{(2)}] Demand that the gauge fields behave like pure gauge by adding Lagrange multipliers $\widetilde Y_{\dm}$
	\begin{equation}
	- \widetilde Y_\dm F_{+ -}^\dm \qq {\rm with } \qq  F_{+ -}^\dm = \partial_+ A_-^\dm - \partial_- A_+^\dm  \ .
	\end{equation}
	
	\item[\textbf{(3)}] Integrate $ A_{\pm} $ out and end up with the dual theory in terms of the dual coordinates $ \widetilde Y_\dm $ and the dual background $ \widetilde Q_d $
	\begin{equation}\label{FacTdualityA}
	\begin{aligned}
	\widetilde Q_{\nii \nj} &=Q_{\nii \nj} - Q_{\nii \dm} (Q^{-1}_d)^{\dm \dn} Q_{\dn \nj} \qq &\widetilde Q^{\dm \dn}& = (Q^{-1}_d)^{\dm \dn}\\
	\widetilde Q_{\nii}{}^{\dn} &= Q_{\nii \dm} (Q^{-1}_d)^{\dm \dn} \qq &\widetilde Q^{\dm}{}_{\nj}& = -(Q^{-1}_d)^{\dm \dn} Q_{\dn \nj}\ .
	\end{aligned}
	\end{equation}

\end{description}

It can also be shown that the transformation of the dilaton is given by \cite{Buscher} (see also \cite{Porrati})
\begin{equation}\label{TdualityPhi}
\widetilde \Phi = \Phi - \frac{1}{2}\ln\left(\text{Det}(Q_d)\right) \ ,
\end{equation}
where $ Q_d $ is the matrix notation for the internal components of $ Q $.

The duality transformations (\ref{FacTdualityA}) can be compared with the general way in which an element $ \psi \in O(d,d) $ acts on the background fields (\ref{dualitycomp}). This requires matching internal indices in both expressions by introducing $ \delta $ matrices to relate tildes with primes
\begin{equation}\label{EprimeEtilde}
Y'{}^{\dm} \equiv \delta^{\dm \dn} \widetilde Y_\dn \qq Q'_{\dm \dn} \equiv \delta_{\dm \dpp} \widetilde{Q}^{\dpp \dq} \delta_{\dq \dn}  \qq \Phi' = \widetilde \Phi  \ ,
\end{equation}
and so
\begin{equation} \label{QprimaAbelian}
Q'_d = \delta \widetilde{Q}_d \delta = \delta Q^{-1}_d \delta = \delta \left(\delta^{-1} Q_d\right)^{-1} \ .
\end{equation}
Just to remind the reader, both coordinates and fields with tildes and primes refer to the dual space. The former carry an unconventional index structure due to the way they are obtained through the Buscher procedure. The latter are defined to coincide componentwise to the former, in such a way that the standard index structure is restored through Kronecker deltas.

After comparison with (\ref{dualitycomp}) and (\ref{DualityPhi}) we can {\it read} the local $O(d,d)\times \mathbb{R}^+$ that connects the original background with its dual
\begin{boxquation}
\begin{equation}\label{Oddabelian}
\psi_\ddm{}^\ddn=\begin{pmatrix}
0 & \delta_{\dm \dn}\\
\delta^{\dm \dn} & 0
\end{pmatrix} \ , \ \ \  \alpha = 0\ .
\end{equation}
\end{boxquation}

This result shows that after Buscher's procedure, we end up with a dual theory obtained by the application of a full factorized transformation (\ref{fullTTduality}) on the background. Had we dualized along a fewer number of isometries, for instance only one, the resulting $O(d,d)$ element would have been a (product of) factorized T-duality (\ref{TTduality}). Let us also mention that while global $GL(d)$ transformations are manifest symmetries of the sigma model, it is also possible to prove that invariance under  $B$-shifts can be achieved by incorporating a boundary term containing a closed 2-form. The combined action of these symmetries spans the {\it full} rigid $O(d,d)$ action on the background.

Because we chose adapted coordinates, the internal twists of the original background $U_M{}^I$ and $\lambda$, and those of the dual background $U'_M{}^I$ and $\lambda'$ are constant and so generate vanishing gaugings (\ref{F})-(\ref{Fprime})
\begin{equation}
    F_{I J K} = F'_{IJK} = 0 \ , \ \ \ \ \ F_I = F'_I = 0 \ .
\end{equation}
In this case it is obvious that both set of gaugings belong to the same duality orbit, and then give rise to the same physics, namely that of an ungauged supergravity.

Since the twist matrix is constant it can be absorbed by $\widehat E$. Then, considering the external Lorentz gauge in which $\widehat e^{(+)} = \widehat e^{(-)}$, we see from (\ref{Ominusgeneral}) that $\Oma_s^{(-)} = 1$ and consequently the Lorentz transformation connecting the supergravity gauge with gauged DFT is trivial.

\subsubsection{Non-Abelian T-duality}\label{sec_NATD}

The non-Abelian counterpart of T-duality \cite{Quevedo} now relies on the target space possessing $ d $ non-commuting isometries $\mathcal{L}_k Q = 0$, generated by a non-Abelian group \go with killing vectors satisfying
\begin{equation*}
\left[k_\da, k_\db\right] = f_{\da \db}{}^{\dc} k_{\dc} \ .
\end{equation*}
To facilitate contact with the discussion at the beginning of this section, this is the first equation in (\ref{leftrightbrackets}). The sigma-model is
\begin{equation}\label{reduced sigmaNA}
S = \int \mathrm{d}^2\sigma \ \left[ \partial_+ X^\nii \partial_- X^\nj Q_{\nii \nj} + R_+{}^\da \partial_- X^\nj Q_{\da \nj} + \partial_+ X^\nii R_-{}^\db Q_{\nii \db} + R_+{}^\da R_-{}^\db Q_{\da \db} \right] \ ,
\end{equation}
where now the internal dependency is encoded in the right-invariant one-forms $ R_m{}^i$, which act as vielbeins exchanging algebraic $ \da,\db=1,\dots,d $ and curved $ \dm,\dn=1,\dots,d $ indices
\begin{equation}\label{NATDcurving}
R_+{}^\da R_+{}^\db Q_{\da \db} = \partial_+ Y^\dm R_\dm{}^\da Q_{\da \db} R_{\dn}{}^\db \partial_- Y^\dn \equiv  \partial_+ Y^\dm Q_{\dm \dn} \partial_- Y^\dn \ .
\end{equation}
This way of writing the background shows that the whole dependence on the internal space is encoded in the Maurer-Cartan forms $R(Y)_m{}^i$, while the components $ Q(X)_{\da \db} $ depend only on the spectator fields (external coordinates). Equation (\ref{NATDcurving}) is useful to note the difference with the Abelian case (\ref{reduced sigma}), where the Maurer-Cartan forms were trivial $R_\dm{}^\da = \delta_\dm{}^\da $ and so using algebraic or curved indices was equivalent.

There is a Buscher-like procedure built by De la Ossa and Quevedo \cite{Quevedo} that leads to an equivalent dual background. The procedure closely follows the one performed before with the difference that the auxiliary fields $ A_{\pm} $ and their strength-energy tensors $ F_{+ -} $ are now valued in a non-Abelian algebra. The result is given by
	\begin{equation}\label{ENA}
	\begin{aligned}
	\widetilde Q_{\nii \nj} &=Q_{\nii \nj} - Q_{\nii \da} \left[ \left( Q_d + f Y'\right){}^{-1}\right]{}^{\da \db} Q_{\db \nj} \qq &\widetilde Q^{\da \db}& = \left[ \left( Q_d + f Y'\right){}^{-1}\right]{}^{\da \db}\\
	\widetilde Q_{\nii}{}^{\db} &= Q_{\nii \da} \left[ \left( Q_d + f Y'\right){}^{-1}\right]{}^{\da \db}
	\qq &\widetilde Q^{\da}{}_{\nj}& = -\left[ \left( Q_d + f Y'\right){}^{-1}\right]{}^{\da \db} Q_{\db \nj} \ ,\\
    \end{aligned}
	\end{equation}
where
\begin{equation*}
(f Y')_{\da \db} \equiv f_{\da \db}{}^\dc \, Y'_\dc \ .
\end{equation*}

Regarding the dilaton field, its transformation can be obtained from \cite{PLTplurality}
\begin{equation}
\widetilde \Phi = \Phi -\frac{1}{2}\ln\left(\text{Det}\left( Q_d + f Y'\right)\right) + \frac{1}{2} \ln \det a\ ,
\end{equation}
 with $a$ defined in (\ref{adLR}), and this reduces to the standard form found by Quevedo and de la Ossa \cite{Quevedo} when the algebra is uni-modular $f_{\da \db}{}^\db = 0$.
To cast this transformation in an $O(d,d)$ format, we first need to express everything in terms of curved indices and then change from the tilde convention to the prime convention as we did in (\ref{EprimeEtilde}). To curve the indices we use the Maurer-Cartan forms of each space. For the original background we rotate with $R_\dm{}^\da$, which connects $Q_{\dm \dn} = R_\dm{}^\da Q_{\da \db} R_\dn{}^\db$, and for brevity we will keep noting this with matrix notation $Q_d$ even though now this is a curved object. The dual background happens to carry an Abelian algebra and so algebraic and curved indices are indistinguishable and related by $\delta_\dm{}^\da$. For instance in the internal sector we have
\begin{equation}
\widetilde Q^{\dm \dn} = \delta_\da{}^\dm \left[ \left( R^{-1} Q_d R^{-t} + f Y'\right){}^{-1}\right]{}^{\da \db} \delta_\db{}^\dn \ .
\end{equation}
 Finally, we lower the indices with $\delta_{\dm \dn}$ as we did in (\ref{EprimeEtilde}) and for brevity we introduce a new mixed-index Kronecker's delta $\delta_{\dm \dn} \delta_{\da}{}^\dn \equiv \delta_{\dm \da} \equiv \delta$. After this procedure, (\ref{ENA}) leads to
\begin{equation}\label{surved_ENA}
\begin{aligned}
Q'_{\nii \nj} &=Q_{\nii \nj} - Q_{\nii \dpp} \left[ \left( Q_d + R f Y' R^t \right){}^{-1}\right]{}^{\dpp \dq} Q_{\dq \nj} \\
Q'_{\nii \dn} &= Q_{\nii \dpp} \left[ \left( \delta^{-1} R^{-1} Q_d + \delta^{-1} f Y' R^t\right){}^{-1}\right]{}^{\dpp}{}_\dn
\\ Q'_{\dm \nj}& = - \delta_{\dm \da} R_\dpp{}^\da \left[ \left( \delta^{-1}R^{-1} Q_d + \delta^{-1}f Y' R^t\right){}^{-1}\right]{}^{\dpp}{}_\dq \delta^{\dq \db} R_\db{}^\dpp Q_{\dpp \nj} \\
Q'_{\dm \dn}& = \delta_{\dm \da} R_\dpp{}^\da \left[ \left( \delta^{-1}R^{-1} Q_d + \delta^{-1}f Y' R^t\right){}^{-1}\right]{}^{\dpp}{}_\dn \ ,
\end{aligned}
\end{equation}
 while for the dilaton we have
\begin{equation}\label{NATDPhi}
\Phi' = \Phi -\frac{1}{2}\ln\left(\text{Det}\left( \delta^{-1} R^{-1}Q_d + \delta^{-1}f Y' R^t\right)\right) + \frac{1}{2}\ln\text{Det}\left(L\right) \ .
\end{equation}

As discussed in (\ref{coordinatedependence}), the RHS of these equations {\it look like} there is a dependence on the original set of coordinates through $R(Y)$, but after some work these equations can be taken to the form
\begin{equation}
\begin{aligned}
Q'_{\nii \nj} &= Q_{\nii \nj} - Q_{\nii \da} \left[ \left( Q_d + f Y' \right){}^{-1}\right]{}^{\da \db} Q_{\db \nj} \\
Q'_{\nii \dn} &= Q_{\nii \da} \left[ \left( Q_d + f Y' \right){}^{-1}\right]{}^{\da \db} \delta_{\db \dn}
\\ Q'_{\dm \nj}& = - \delta_{\dm \da} \left[ \left( Q_d + f Y'\right){}^{-1}\right]{}^{\da \db} Q_{\db \nj} \\
Q'_{\dm \dn}& = \delta_{\dm \da} \left[ \left( Q_d + f Y'\right){}^{-1}\right]{}^{\da \db} \delta_{\db \dn} \\
\Phi' &= \widehat \Phi -\frac{1}{2}\ln\left(\text{Det}\left(Q_d + f Y' \right)\right) \ ,
\end{aligned}
\end{equation}
where $Q_d \equiv Q(X)_{\da \db}$ and we are considering a non-trivial background for the dilaton $\Phi(X,Y) = \widehat \Phi(X) - \frac{1}{2}\ln \det{ a (Y)}$, which is isometric except in the non-unimodular case $\mathcal{L}_{k_i} \Phi = \frac{1}{2} f_{i j}{}^j$. It is then clear that the dual fields depend only on the dual coordinates $Y'$ only.

The expressions (\ref{surved_ENA}) and (\ref{NATDPhi}) can now be compared directly with (\ref{dualitycomp}) and (\ref{DualityPhi}) to recognize the $O(d,d) \times \mathbb{R}^+$ transformation that connects the dual backgrounds
\begin{boxquation}
\begin{equation}\label{NATDmatrixcurved}
\psi(Y, Y')_\ddm{}^\ddn = \begin{pmatrix}
0 & \delta_{\dm \db} R(Y)_\dn{}^\db\\
\delta^{\dm \db} R(Y)_\db{}^\dn & \delta^{\dm \dc} f_{\dc \db}{}^\dd Y'_\dd R(Y)_\dn{}^\db
\end{pmatrix} \ , \ \ \ \ \alpha(Y) = \frac{1}{2}\ln\text{Det}\left(L(Y)\right) \ .
\end{equation}
\end{boxquation}
We see immediately  that (\ref{NATDmatrixcurved}) reduces to the Abelian case (\ref{Oddabelian}) when $f_{\da \db}{}^\dc = 0$ and $R_\dm{}^\da = L_\dm{}^\da = \delta_{\dm}{}^\da $.

Let us briefly discuss what happened above in the language of Gauged DFT. We started with the original generalized background
\begin{equation} \label{UlambdaNATDorig}
    U(Y)_\ddm{}^\dda =
    \begin{pmatrix}
    R(Y)_\dm{}^\da & 0\\
    0 & R(Y)^\dm{}_\da
    \end{pmatrix} \ , \ \ \ \  \lambda(Y) = - \frac{1}{2} \ln \text{Det}\left(L(Y)\right) \ ,
\end{equation}
corresponding to a geometric background with vielbein $R_m{}^i$, dilaton background $-\frac{1}{2}\ln \det a $ and vanishing 2-form flux. As such, the only components of the gaugings $F_{\dda \ddb \ddc}$ and $F_\dda$ are given by metric fluxes (\ref{F})
\begin{equation}\label{FluxesNATDOrig}
   F_{\da \db \dc} = F_\da{}^{\db \dc} = F^{\da \db \dc} = 0 \ ,  \ \ \ F_{\da \db}{}^\dc = - f_{\da \db}{}^\dc \ , \ \ \ F_\dda = 0 \ .
\end{equation}
These identities follow from the Lie bracket of $R^{-1}$ (\ref{leftrightbrackets}) and the Jacobi identity. After the dualization, we ended with a different generalized background
\begin{equation}\label{dualNATD}
    U'(Y'){}_\ddm{}^\dda = \begin{pmatrix}
    0 & \delta_{\dm \da}\\
    \delta^{\dm \da} & \delta^{\dm \db} \left(f Y'\right)_{\db \da}
    \end{pmatrix} \ , \ \ \ \  \lambda'(Y') = 0 \ ,
\end{equation}
that yields the exact same gaugings (\ref{Fprime}) except for the vectorial flux $F_\dda$ which picks up a contribution from the trace of the structure constants
\begin{equation} \label{FluxesNATDDual}
    F'_{\da \db \dc} = F'_\da{}^{\db \dc} = F'{}^{\da \db \dc} = 0 \ ,  \ \ \ F'_{\da \db}{}^\dc = - f_{\da \db}{}^\dc \ , \ \ \ F'_i = f_{\da \db}{}^\db \ , \ \ \ F'{}^i=0 \ .
\end{equation}
In can be checked that (\ref{FluxesNATDOrig}) and (\ref{FluxesNATDDual}) satisfy the consistency conditions (\ref{quadraticconstraints}).

We now discuss two distinct cases. If the group were unimodular $f_{ij}{}^j = 0$, then both set of gaugings (\ref{FluxesNATDOrig}) and (\ref{FluxesNATDDual}) would coincide exactly.  As a consequence, the Gauged DFT would remain invariant under the local $ O(d,d)\times \mathbb{R}^+$ transformation (\ref{NATDmatrixcurved}) yielding the physical equivalence of both backgrounds, at least at the classical level. It would have been enough that both gaugings fell into the same orbit, but interestingly in this case they happen to coincide. Instead, if the group is {\it not} unimodular $f_{ij}{}^j \neq 0$, then both set of gaugings (\ref{FluxesNATDOrig}) and (\ref{FluxesNATDDual}) fall into {\it different} duality orbits, and we loose guaranty that if the original background is a solution to the DFT equations of motion, so is the dual background. Note however that if the dual background ($U'$, $\lambda'$) in (\ref{dualNATD}) is {\it deformed} into ($U'$, $\lambda' + \widetilde \lambda'$), with
\begin{equation} \label{lambdatilde}
    \widetilde \lambda' = - \frac 1 2 f_{i j}{}^j \delta^{m i} \widetilde Y'_m \ ,
\end{equation}
the gaugings of this deformed background coincide with those of the original background
 \begin{equation} \label{FluxesDualesFinal}
    F'_{\da \db \dc} = F'_\da{}^{\db \dc} = F'{}^{\da \db \dc} = 0 \ ,  \ \ \ F'_{\da \db}{}^\dc = - f_{\da \db}{}^\dc \ , \ \ \ F'_I = 0 \ .
\end{equation}
 Then, this background is indeed a solution to the equations of motion of DFT. We can interpret this fact as follows. The deformed background is a composition of two successive reductions: one with twist ($1$, $\widetilde \lambda'$) and another one with twist ($U'$, $\lambda'$). The first twist ($1$, $\widetilde \lambda'$) produces a first gauging of DFT with fluxes
 \begin{equation} \label{GaugedIntermediate}
     F'_{I J K} = 0 \ , \ \ \ F'_i = - f_{\da \db}{}^\db \ , \ \ \ F'{}^i=0 \ .
 \end{equation}
 The second twist reduces this Gauged DFT into another one with gaugings (\ref{FluxesDualesFinal}), which now happily fall into the same duality orbit than (\ref{FluxesNATDOrig}). Then, the local $O(d,d) \times \mathbb{R}^+$ (\ref{NATDmatrixcurved}) maps a solution (\ref{UlambdaNATDorig}) of ungauged DFT, to a solution (\ref{dualNATD}) of a Gauged DFT with gaugings (\ref{GaugedIntermediate}). Interestingly, the gauging (\ref{GaugedIntermediate}) leads to the deformations of the DFT equation of motions, which on section happen to correspond to the so-called generalized supergravity equations \cite{GSE}, as discussed in \cite{DoubleAspects}, \cite{GSEfromDFT}, \cite{Baguet:2016prz}.

Interestingly, since the original space is a geometric background (\ref{UlambdaNATDorig}) we see that, understood as a particular case of (\ref{Egeneral}), namely $u = R$, $b = \beta = 0$, the double Lorentz transformation $L_s$ in Figure \ref{DFTdiagram} is trivial in the external gauge $\widehat e^{(+)} = \widehat e^{(-)} $ which induces $ e^{(+)} = e^{(-)} $. We then have $\Oma_s^{(-)} = 1$, which means that the supergravity gauge is already in the gSS form.

The question remains on how to interpret the dual background (\ref{dualNATD}) in the context of Gauged DFT. It is difficult to {\it read} a background from a generalized twist due to the double Lorentz symmetry. To avoid this ambiguity, it is instructive to build the generalized metric for the dual background
\begin{equation}\label{Genparalelization}
    {\cal H}'_{\ddm \ddn}  = U'{}_\ddm{}^\dda \delta_{\dda \ddb} U'{}_\ddn{}^\ddb =
\begin{pmatrix}
\delta_{\dm \dn} & \delta_{\dm p}\beta^{p n} \\
- \beta^{\dm p} \delta_{p n} & \delta^{\dm \dn} - \beta^{\dm p} \delta_{p q} \beta^{q \dn}
\end{pmatrix} \ \ \  {\rm where } \ \ \ \beta^{m n} = - \delta^{\dm \da} (f Y')_{\da \db} \delta^{\db \dn}  \ ,
\end{equation}
where we set the scalar fluctuations to zero ${\cal H}_{\dda \ddb} = \delta_{\dda \ddb}$. This form of the parameterization in terms of a bi-vector is typical of globally non geometric backgrounds (see for example \cite{exploringDFT}, \cite{NonGeom}). It is clear from here that the background is locally geometric, but globally it corresponds to the wired case of a non-geometric background with a generalized paralellization that renders the gaugings purely geometric. Let us explain this a little further. The background is usually simple to read from the background generalized metric, while reading it from the twist matrix is cumbersome due to the redundancy produced by the  choice of the internal double Lorentz gauge. This choice fixes the generalized paralellization \cite{GenParal}. It doesn't affect the background, but it does change the fluxes and then has a crucial impact on the lower dimensional physics. The paradigmatic case is that of a torus parallelized in a funny way that yields the fluxes of a sphere \cite{DualityOrbits} (see also \cite{Enhancement}).

\subsubsection{Yang-Baxter deformations}\label{sec_YB}

Yang-Baxter deformations \cite{KlimcikYB} relate backgrounds associated to integrable systems \cite{KlimcikYBint}. They are based on an $ \text{R}^{\da \db}$-matrix (not to be confused with the right-invariant one-form $R_\dm{}^\da$) satisfying the algebraic equation
\begin{equation}\label{CYBE}
\left[\text{R}X,\text{R}Y\right] - \text{R}\left(\left[\text{R}X, Y\right] + \left[X,\text{R}Y\right]\right)= c^{2}\left[X,Y\right] \ ,
\end{equation}
where $c \in [-1,0,1]$, $ X,Y \in \ga$  and $ \left[ \ , \ \right] $ is the Lie-bracket of the isometry algebra of the background to be deformed. The case $ c=0 $ corresponds to classical YB equations (CYBE) (also called homogeneous equations) and $ c \neq 0 $ leads to so-called modified classical YB equations. The latter cases leads to in-homogeneous YB deformations, sometimes called $\eta$-deformations, and they have been widely study in the context of AdS$_5 \times $ S$_5$ backgrounds\cite{Delduc}. Here we will concentrate on CYBE only, which lead to homogeneous YB deformations, because its connection to NATD is simpler. These transformations preserve conformal invariance if the $ \text{R}$-matrix is unimodular \cite{UnimodularYB}
\begin{equation}\label{unimodularcondition}
\text{R}^{\da \db} f_{\da \db}{}^\dc = 0 \ .
\end{equation}
It was conjectured in \cite{NATDYB} that the homogeneous Yang-Baxter model can be obtained by applying NATD to the original background, with respect to an isometry group determined by the $ \text{R}$-matrix. This conjecture was proven in \cite{NATDYBWB} and \cite{NATDYBWB2} for principal chiral models where rules were established for connecting NATD and YB models.

Picturing YB deformations as NATDs  requires a {\it dressed} $\text{R}$ operator
\begin{equation}
\text{R}_g \equiv \text{Ad}_{g^{-1}} \text{R} \text{Ad}_{g} \ , \ \ \ \ (\text{R}_g)^{\da \db} = a_\dc{}^\da \text{R}^{\dc \dd} a_\dd{}^\db = R_\dm{}^\da k_\dc{}^\dm \text{R}^{\dc \dd} k_{\dd}{}^\dn R_\dn{}^\db \ ,
\end{equation}
 and identifying
\begin{equation}\label{YBNATDrule}
f_{\da \db}{}^\dc Y'_{\dc} = \eta^{-1} (\text{R}^{-1})_{\da \db} - \eta^{-1} (\text{R}^{-1}_{g})_{\da \db} \ , \ \ \  \ d Y'_\da = \eta^{-1} (\text{R}_g^{-1})_{\da \db} R^\db \ ,
\end{equation}
in the NATD background (\ref{ENA}), where $ \eta $ is called the deformation parameter.

In \cite{BW0}  the NATD transformations for the Green-Schwarz (GS) superstring with a generic isometry group were derived. Using the rules between NATD and YB, the authors also deduced the form of homogeneous YB deformations for a generic GS sigma model given by \cite{eoin}
\begin{equation}\label{YB}
Q'_d = Q_d\left(\eta \Theta Q_d + 1\right) \ , \ \ \  \Phi' = \Phi - \frac{1}{2}\ln\left(\text{Det}(\eta \Theta Q_d + 1)\right) \ ,
\end{equation}
where
\begin{equation}
\Theta^{\dm \dn} \equiv k_\da{}^\dm \text{R}^{\da \db} k_\db{}^\dn =  R_i{}^m \text{R}_g^{i j} R_j{}^n\ ,
\end{equation}
is nothing but the curved version of the dressed R operator.
Using the killing equations and closure of the algebra, the CYBE (\ref{CYBE}) translates into
\begin{equation}\label{YBeq}
\Theta^{\dq [\dm} \partial_{\dq} \Theta^{\dn \dpp]} = 0 \ .
\end{equation}

Comparing (\ref{YB}) with the general formulas (\ref{dualitycomp}) and (\ref{DualityPhi}) one can identify the YB transformation with the following local $ O(d,d) \times \mathbb{R}^+$ transformation  \cite{TurkeB}
\begin{boxquation}
	\begin{equation}\label{YBQ2}
	\begin{aligned}
	\psi(Y)_\ddm{}^\ddn = \begin{pmatrix}
	\delta_\dm{}^\dn & 0\\
	\eta \Theta(Y)^{\dm \dn} & \delta^\dm{}_\dn
	\end{pmatrix} \ , \ \ \  \alpha = 0 \ .
	\end{aligned}
	\end{equation}
\end{boxquation}
The original and deformed backgrounds depend on the same set of coordinates $Y$. The interpretation of YB as the non-Abelian extension of $ \beta$-shifts\cite{YBandbetaabelian}-\cite{SakatanibetaYB},\cite{Lust} (also known as TsT transformations) can be easily seen from here. When $ \text{R} $ is defined in an Abelian sub-algebra of the isometry algebra, the killing vectors and Maurer-Cartan forms are trivial $ k=R= 1 $ and so $ \Theta $ and $\text{R}$ are constant. As a consequence (\ref{YBQ2}) reduces exactly to a constant beta shift (\ref{betaTduality}).

We can check if the fluxes generated by the original and dual background indeed fall into the same duality orbit. Consider the original background described in Gauged DFT by generic twists  $U(Y)_\ddm{}^\dda$ and $\lambda(Y)$ depending only on the supergravity coordinates. The deformed background is defined over the same set of coordinates $U'(Y)_\ddm{}^\dda = \psi(Y)_\ddm{}^{\ddn} U(Y)_\ddn{}^\dda$  and $\lambda'(Y) = \lambda(Y) + \alpha = \lambda(Y)$. It can be shown that the isometric condition for the background fields and the uni-modularity condition ensure that both $F_{\dda \ddb \ddc}$ and $F_{\dda}$ remain invariant\cite{Lust},\cite{BW2},\cite{Bakhmatov:2018bvp}. This can be seen by splitting
\begin{equation}
\Psi_\ddm{}^\ddn = \delta_\ddm{}^\ddn + \eta \Theta_\ddm{}^\ddn \ , \ \ \
\Theta^{\ddm \ddn} = k_{\da}{}^\ddm \text{R}^{\da \db} k_\db{}^\ddn \equiv
\begin{pmatrix}
\Theta^{\dm \dn} & 0\\
0 & 0
\end{pmatrix} \ , \ \ \  k_{\da}{}^\ddm \equiv \left(k_{\da}{}^\dm, 0\right) \ ,
\end{equation}
which gives
\begin{eqnarray}
F'_{\dda \ddb \ddc} &=& F_{\dda \ddb \ddc} + 3 \eta \left(U_{\ddp[\dda } k_{i}{}^\ddp \text{R}^{i j} U_{\ddn\ddc } \widehat{\mathcal{L}}_{k_j} U^\ddn{}_{\ddb]} +  \eta \Theta^{\ddm [\ddp} \partial_\ddm  \Theta^{\ddq \ddn]} U_{\ddp\dda } U_{\ddq\ddb } U_{\ddn\ddc }\right) = F_{\dda \ddb \ddc} \label{FIJK2}\\
F'_{\dda} &=& F_{\dda} + 2 \eta \left( U_{\ddp\dda } k_\da{}^\ddp \text{R}^{\da \db} \widehat{\mathcal{L}}_{k_\db} \lambda + \text{R}^{\dd \de} k_\dd{}^\ddp \partial_\ddp k_\de{}^\ddm U_{\ddm\dda } \right)  \ , \label{FI2}
\end{eqnarray}
where $\widehat{\mathcal{L}}_{k_\da}$ is the generalized Lie derivative. If the twist $U$ is generalized isometric with respect to the generalized killing vector $k_i$, then the first term in (\ref{FIJK2}) vanishes, while the second vanishes due to the YB equation (\ref{YBeq}). Then, the fluxes $F_{I J K}$ remain invariant. Regarding the fluxes $F_I$, the first term in  (\ref{FI2}) vanishes if the twist $\lambda$ is generalized isometric, while the second term vanishes if the group is unimodular (\ref{unimodularcondition}). If this is the case, then the dual gaugings fall into the same duality orbit. If not, a procedure similar to (\ref{lambdatilde}) is required in order to interpret the dual background as a solution to a deformed theory. Note however that in this case, the dual vectorial fluxes would be non-constant, and so it is unclear to us if they can be generated through a twist in Gauged DFT. A similar discussion on this point will take place in PL T-duality.

Finally it is worth mentioning that, in analogy with the NATD case, if we consider the original space as a geometric background the double Lorentz transformation $L_s$ is again trivial $\Oma_s^{(-)} = 1$.

\subsubsection{Poisson-Lie T-duality}

In \cite{Klimcik1}, Klimcik and Severa brilliantly abandoned the requirement of isometries  as the guiding principle for duality, replacing it by a higher algebraic structure that relates dual models, in which isometries only show up in special cases. We will review the procedure restricting attention to the internal sector, so the  expectator fields will be frozen. The starting point is then the internal sector of a generic sigma-model
\begin{equation}\label{PLS0}
S = \int d^2 \sigma \ \partial_+ Y^\dm \partial_-Y^\dn Q_{\dm \dn}  \ ,
\end{equation}
where the group \go acts freely and transitively. It transforms the coordinates as in (\ref{coordinate_change}) $\delta Y^\dm = \epsilon^\da(\sigma_\pm) k_\da{}^\dm{} $, inducing the following change in the action
\begin{equation}\label{generalSvariation}
\delta S = \int d^2\sigma \epsilon^\da \left[\partial_+ Y^\dm \partial_- Y^\dn \mathcal{L}_{k_\da} Q_{\dm \dn}\right] - \int \left[\epsilon^\da d J_\da + d \left(\epsilon^\da J_\da\right)\right] \ ,
\end{equation}
where we defined the Noether currents
\begin{equation}
J_\da \equiv k_{\da}{}^\dm \left(Q_{\dm \dn} \partial_- Y^\dn d \sigma^- - Q_{\dn \dm} \partial_+ Y^\dn d \sigma^+\right) \ .
\end{equation}

Neglecting the global term in (\ref{generalSvariation}), the Abelian and non-Abelian T-duality scenarios are recovered by considering \go as the isometry group of the target space in which $k$ are the killing vectors. The interesting point is that the invariance of the action can still be satisfied without isometries. The idea is to  think of $ J_\da $ as the components of an element $ J $ of a dual algebra $ \tga $
\begin{equation}
J = J_\da t'{}^\da \qq [t'{}^\da, t'{}^{\db} ] = f'{}^{\da \db}{}_\dc t'{}^\dc \ ,
\end{equation}
with an associated Maurer-Cartan equation
\begin{equation}\label{maurercartan}
d J_\da = \frac{1}{2} f'{}^{\db \dc}{}_\da J_\db \wedge J_\dc \ .
\end{equation}

The invariance of the action, namely the vanishing of (\ref{generalSvariation}), leads to a non-isometric condition on the background
\begin{equation}\label{PLcondition}
\mathcal{L}_{k_\da} Q_{\dm \dn} = - {f'}{}^{\db \dc}{}_\da k_{\db}{}^{\dpp} k_\dc{}^{\dq} Q_{\dm \dpp} Q_{\dq \dn} \ ,
\end{equation}
and analyzing the closure of the algebra over it  leads to a bi-algebraic condition \cite{Drinfeld}, \cite{bialgebras}
\begin{equation}\label{bialgebracondition}
f_{\da \db}{}^\de {f'}{}^{\dc \dd}{}_\de = 4 f_{\de [\da}{}^{[\dc} {f'}{}^{\dd] \de}{}_{\db]} \ ,
\end{equation}
which can be conceived as the mixed components of the Jacobi identities of an extended algebra
\begin{equation}\label{bialgebrabrackets}
\left[t_\da,t_{\db}\right]= f_{\da \db}{}^\dc t_\dc \ , \ \ \ \left[{t}'{}^\da,{t}'{}^{\db}\right]= {f}'{}^{\da \db}{}_\dc {t}'{}^\dc \ , \ \ \ \left[t_\da, {t}'{}^{\db}\right] = f_{\dc \da}{}^{\db} {t}'{}^{\dc} - {f}'{}^{\dc \db}{}_{\da} t_\dc \ .
\end{equation}

To enforce that both algebras appear on an equal footing in this framework, a dual  background $Q'_{\dm \dn}$ is introduced
\begin{equation}\label{PLconditiondual}
\mathcal{L}_{{k'}{}^\da} {Q'}{}_{\dm \dn} = - f_{\db \dc}{}^{\da} {k'}{}^{\db \dpp} {k'}{}^{\dc \dq } {Q'}{}_{\dm \dpp} {Q'}{}_{\dq \dn}  \ ,
\end{equation}
together with a dual version of the algebraic identities (\ref{adLR})-(\ref{leftrightbrackets})
\begin{equation}\label{MaurerCartanpropdual}
\begin{aligned}
\text{Ad}_{g'{}^{-1}} t'{}^{\da} &= g'{}^{-1} t'{}^{\da} g' = a'{}^{\da}{}_\db t'{}^{\db} \ ,  \ \ \ {L'}{}_{\dm \da} = \left({g'}{}^{-1}{\partial'}_\dm {g'}\right){}_{\da} \ , \ \ \  {R'}{}_{\dm \da} = \left({\partial'}_\dm {g'} g'{}^{-1}\right){}_{\da} \ ,\\
{k}'{}^{\da \dm} {L'}_{\dm \db} &= \delta^\da{}_{\db} \ , \ \ \  k'{}^{\da \dm} R'_{\dm \db} = (a'{}^{-1})^\da{}_\db \ , \ \ \  R'_{\dm \db} a'{}^{\db}{}_\da = L'_{\dm \da} \ ,
\end{aligned}
\end{equation}
which defines dual left and right invariant Maurer-Cartan forms $L'_{mi}, R'_{m i}$. Now we have a dual group \got  and $ {\partial'}_\dm \equiv \frac{\partial}{\partial {Y}'{}^\dm} $ with $ {Y}'{}^\dm $ the coordinates of the dual manifold. Everything is now doubled, and starts to smell like DFT.

Combining the bi-algebraic condition (\ref{bialgebracondition}) with the introduction of a non-degenerate, ad-invariant bilinear form $ \biform $ satisfying
\begin{equation}\label{Maxiso}
\biform[t_\da][t_{\db}] = \biform[{t'}{}^\da][{t'}{}^{\db}] = 0 \ , \ \ \  \biform[t_\da][{t'}{}^{\db}] = \delta_{\da}{}^{\db} \ ,
\end{equation}
one can identify $ \ga $ and $ \tga $ with the maximally isotropic subalgebras of a Drinfeld double $ \mathcal{D} $\cite{Drinfeld}. It was shown in \cite{Klimcik1} and \cite{PLcanonical} that the sigma-models associated to $ Q $ and $ Q' $ are related by a canonical transformation, so both backgrounds satisfy the same equations of motion.

Using the structure of Drinfeld doubles one can build solutions to the PL conditions (\ref{PLcondition}) and (\ref{PLconditiondual}) given by \cite{Klimcik1}
\begin{equation}\label{PLsolution}
\begin{aligned}
Q_{\dm \dn} = R_\dm{}^\da [\cero Q^{-1} - \pi ]^{-1}_{\da \db} R_\dn{}^{\db} \ , \ \ \ \  {Q'}_{\dm \dn} = {R'}_{\dm \da} [\cero Q - {\pi'} ]^{-1}{}^{\da \db} {R'}_{\dn \db} \ ,
\end{aligned}
\end{equation}
where
\begin{equation}
\pi^{\da \db} \equiv c^{\da \dc} (a^{-1})_\dc{}^\db = - \pi^{\db \da} \ , \ \ \ \ \pi'_{\da \db} \equiv c'_{\da \dc} (a'{}^{-1})^\dc{}_\db = - \pi'_{\db \da} \ ,
\end{equation}
and the matrices $a(g)$, $c(g)$, $a'(g')$ and $c'(g')$  are defined by the adjoint action
\begin{equation}\label{Adg}
\text{Ad}_{g^{-1}}(t_\da) = a_{\da}{}^{\db}\, t_{\db} \ , \ \ \  \text{Ad}_{g^{-1}}({t'}^\da) = c^{\da \db} t_{\db} + (a^{-t})^\da{}_\db {t'}^{\db}\ ,
\end{equation}
while for the dual matrices we have
\begin{equation}\label{Adgdual}
\text{Ad}_{g'{}^{-1}}(t'{}^{\da}) = a'{}^{\da}{}_{\db}\, t'{}^{\db} \ , \ \ \  \text{Ad}_{g'{}^{-1}}({t}_\da) = c'_{\da \db} t'{}^{\db} + (a'{}^{-t})_\da{}^\db {t}_{\db}\ .
\end{equation}
Regarding the field $\cero Q$, it is a constant matrix that would depend on external coordinates if the spectator fields were taken into account. It comes from the construction of the explicit solutions (\ref{PLsolution}) to the PL conditions (\ref{PLcondition}) and (\ref{PLconditiondual}), corresponding to the original background evaluated at the identity of the Drinfeld double, i.e. $\cero Q = Q(e) $.

The expression for the dilatons were originally given in \cite{PLdilaton} and latter improved in \cite{PLTplurality} in the context of PL-plurality (see also \cite{Sakatani})
\begin{equation}\label{PLPhi}
\Phi = \cero \Phi - \frac{1}{2}\ln\text{Det}\left(\frac{\cero Q}{R^{-1} Q_d R^{-t} a^{-t}}\right) \ , \ \ \ \ \Phi' = \cero \Phi - \frac{1}{2}\ln\text{Det}\left(\frac{1}{R'{}^{-1} Q_d' R'{}^{-t} a'{}^{-t}}\right) \ ,
\end{equation}
where, $\cero \Phi$ can be taken to be a constant, that on general grounds would depend only on the expectator fields\footnote{PL-duality works even if $\cero \Phi$ depends on the internal coordinates \cite{PLTplurality}.}.

Elimination of $\widehat Q$ and $ \widehat \Phi$ in  (\ref{PLsolution}) and (\ref{PLPhi}) leads to
\begin{equation}\label{PLmeasure}
\begin{aligned}
Q'_d &= \left(R' \pi R^{-1} Q_d + R' R^t\right)\left[\left(R'{}^{-t} R^{-1} - R'{}^{-t} \pi' \pi R^{-1}\right) Q_d  - R'{}^{-t}\pi' R^t \right]^{-1} \\
\Phi' &= \Phi - \frac{1}{2}\ln \text{Det}\left[\left(R'{}^{-t} R^{-1} - R'{}^{-t} \pi' \pi R^{-1}\right) Q_d - R'{}^{-t}\pi' R^t  \right] + \frac{1}{2}\ln \text{Det}(L) - \frac{1}{2}\ln \text{Det}(L') \ .
\end{aligned}
\end{equation}
Notice once again that although here it {\it looks like} the RHS depends on the original set of coordinates $Y$ through $L(Y)$, $R(Y)$ and $\pi(Y)$, in reality they only depend on $Y'$ as is clear from (\ref{PLsolution}) and (\ref{PLPhi}). These expressions (\ref{PLmeasure}) can now be compared with (\ref{dualitycomp}) and (\ref{DualityPhi}) to recognize the $O(d,d) \times \mathbb{R}^+$ transformation that connects the dual backgrounds
\begin{boxquation}
\begin{equation} \label{OddRPL}
\psi(Y,Y')_\ddm{}^\ddn = \begin{pmatrix}
R' \pi R^{-1} & R' R^t\\
R'{}^{-t} R^{-1} - R'{}^{-t} \pi' \pi R^{-1} & - R'{}^{-t}\pi' R^t
\end{pmatrix} \ , \qq \alpha(Y,Y') = \frac{1}{2}\ln \text{Det}(L) - \frac{1}{2}\ln \text{Det}(L') \ ,
\end{equation}
\end{boxquation}
where the unprimed  components depend on $Y$, and the primed ones on $Y'$.

From the point of view of Gauged DFT, the solutions (\ref{PLsolution}) and (\ref{PLPhi}) can be interpreted in terms of a gSS ansatz (\ref{SSansaetz}) in which $\cero Q$ and $\cero \Phi$ are the external coordinate dependent fields encoded in $\cero \Eext(X)$ and $\cero d(X)$, the twists of the original background are given by
\begin{equation}\label{PLU}
U(Y)_M{}^I = \begin{pmatrix}
R(Y) & 0\\
-R(Y)^{-t}\pi(Y) & R(Y)^{-t}
\end{pmatrix} \ , \ \ \ \lambda(Y) = -\frac{1}{2}\ln \text{Det}(L(Y)) \ ,
\end{equation}
and those of the dual background by
\begin{equation}\label{PLUdual}
U'(Y') = \begin{pmatrix}
0 & R'(Y')\\
R'(Y'){}^{-t} & -R'(Y'){}^{-t} \pi'(Y')\end{pmatrix} \ , \ \ \ \lambda'(Y') = -\frac{1}{2}\ln \text{Det}(L'(Y')) \ .
\end{equation}
Both backgrounds are connected by the local $O(d,d) \times \mathbb{R}^+$ transformation (\ref{OddRPL}) as $U'(Y') = \psi(Y,Y') U(Y)$ and $\lambda'(Y') = \lambda(Y) + \alpha(Y,Y')$. It is then clear from (\ref{ExternalCorr1}) and (\ref{ExternalCorr2}) that higher derivatives enter the solutions {\it only} though $\widehat Q$ and $\widehat \Phi$.

Before we compute the gaugings, let us show how the previously introduced expressions can be cast into a {\it double} language (see for example \cite{DoubleAspects}). Grouping the generators into a double generator $T_\dda = (t_\da, t'{}^{\da})$ permits to cast the maximal isotropic condition (\ref{Maxiso}) in terms of the $O(d,d)$ invariant matrix
\begin{equation}\label{isotropicODD}
\biform[T_\dda][T_\ddb] = \eta_{\dda \ddb} = \begin{pmatrix}
0 & \delta_\da{}^\db\\
\delta^\da{}_\db & 0
\end{pmatrix} \ ,
\end{equation}
and also regroup the algebra (\ref{bialgebrabrackets}) in an $O(d,d)$ covariant fashion
\begin{equation}
\left[T_\dda, T_{\ddb}\right] = -F_{\dda \ddb}{}^\ddc T_\ddc \ , \ \ \  F_{\da \db}{}^\dc = - f_{\da \db}{}^\dc \ , \ \ \ \ F{}^{\da \db}{}_\dc = - f'{}^{\da \db}{}_\dc \ .
\end{equation}
The ad-invariant condition over $ \biform $ can then be written as
\begin{equation}\label{Adinvariant1}
\langle \left[T_\dda,T_{\ddb}\right] , T_\ddc \rangle = \langle T_{\ddb}, \left[T_\ddc,T_\dda\right] \rangle \ , \ \ \ \biform[T_\dda][T_{\ddb}] = \biform[g T_\dda g^{-1}][ g T_{\ddb} g^{-1}]   \ .
\end{equation}
Of course we will see that these generalized structure constants $F_{\dda \ddb}{}^\ddc$ are exactly the gaugings generated by both backgrounds. We finally point out that the adjoint actions (\ref{Adg}) and (\ref{Adgdual}) can be also combined into an $O(d,d)$ form
\begin{equation}
(\text{Ad}_{g})_\dda{}^\ddb =
\begin{pmatrix}
(a^{-1})_\da{}^\db &  0\\
(c^t)^{\da \db} & (a^{t})^\da{}_\db
\end{pmatrix} \ ,  \ \ \ \  (\text{Ad}_{g'})_\dda{}^\ddb =
\begin{pmatrix}
(a'{}^{t})_\da{}^\db &  c'{}^{t}_{\da \db}\\
0 & (a'{}^{-1})^{\da}{}_\db
\end{pmatrix} \ ,
\end{equation}
where we read $\text{Ad}_{g^{-1}}$ and $\text{Ad}_{g'{}^{-1}}$ from (\ref{Adg}) and (\ref{Adgdual}) and then inverted the matrices. These matrices can be contracted with double left-invariant 1-forms
\begin{equation}
\mathbb{L}_\ddm{}^\dda =
\begin{pmatrix}
L_\dm{}^\da & 0\\
0 & L^\dm{}_\da
\end{pmatrix} \ , \ \ \  \mathbb{L'}_{\ddm \dda} =
\begin{pmatrix}
L'_{\dm \da} & 0\\
0 & L'{}^{\dm \da}
\end{pmatrix}  \ ,
\end{equation}
in order to obtain the twist matrices (\ref{PLU}) and (\ref{PLUdual})
\begin{equation}
U_\ddm{}^\dda  = \mathbb{L}_\ddm{}^\ddb{} (\text{Ad}_g)_\ddb{}^\dda \ , \ \ \ U'_\ddm{}^\dda  = \mathbb{L}'_{\ddm}{}^\ddb (\text{Ad}_{g'})_\ddb{}^\dda \ .
\end{equation}

Having written everything in double language, it is now obvious that we can rotate every object carrying indices $I,J,K,\dots$ with rigid elements $ h \in O(d,d)$, which is simply a renaming that does not change the results. In the language of Gauged DFT this simply amounts to translations withing a fixed duality orbit, as discussed around (\ref{fluxtransf}). In the context of generalized dualities, these rotations are known as PL T-pluralities \cite{PLTplurality}. This is a generalization of PL T-duality which considers that a Drinfeld double $\mathcal{D}$, can be decomposed in several maximally isotropic subalgebras $\ga$ and $\tga$. Together with the Lie algebra of the Drinfeld $\dalg$, every such decomposition $ \left(\dalg,\ga,\tga\right) $ is known as a Manin triple $ \mani(\mathcal{D}) $.  An important remark is that for any $ \mathcal{D} $ at least we have two Manin triples $ \left(\dalg,\ga,\tga\right) $ and $ \left(\dalg,\tga,\ga\right) $, connected by a full factorized $O(d,d)$ rotation, which from the point of view of the bialgebra are distinct objects. Any such decomposition will give rise to a different background but all of them will be dual to each other. In this scenario, all models are connected by rigid $O(d,d)$ rotations preserving the bi-algebra (\ref{bialgebrabrackets}) and the maximally isotropic condition (\ref{isotropicODD})
\begin{equation}
T'_\dda = h_\dda{}^\ddb T_\ddb \ , \ \ \ \biform[T'_\dda][T'_\ddb] = \eta_{\dda \ddb} \ .
\end{equation}

We can finally compute the gaugings in the context of Gauged DFT defined by the twists (\ref{PLU}) and (\ref{PLUdual}), yielding
\footnote{ To facilitate the computation of the fluxes, we list some useful identities (see also the appendix of \cite{PLcanonical}). The ad-invariance condition of the bilinear form (\ref{Adinvariant1}) implies
\begin{equation}\label{PL_identities}
\begin{aligned}
& a_{\db}{}^\dd a_\dc{}^\de (a^{-1})_\df{}^\da f_{\dd \de}{}^\df  = f_{\db \dc}{}^\da \ , \qq \qq \qq \qq \qq \qq \qq \qq (a^{-1})_{\dd}{}^{\db} (a^{-1})_{\de}{}^\dc a_{\da}{}^\df {f'}^{\dd \de}{}_\df  - {f'}^{\db \dc}{}_{\da} = 2 f_{\da \dd}{}^{[\db} \pi^{\dc]\dd} \\
& (a^{-1})_\dd{}^\db (a^{-1})_\de{}^\dc f'{}^{\dd \de}{}_\df c^{\da \df} = f_{\dd \de}{}^\da \pi^{\dd \db} \pi^{\de \dc} - 2 f'{}^{\da[\db}{}_\dd \pi^{\dc]\dd} \ , \ \ \ \ \ \ \ \ \  \ \ \ \  f'{}^{[\da \db}{}_\dd \pi^{\dc] \dd} - \pi^{\dd [\da} f_{\dd \de}{}^\db \pi^{\dc] \de} = 0\ .
\end{aligned}
\end{equation}
Analogous identities can be obtained for the dual objects by just adding/removing  primes and exchanging the position of all indices.
We finally point out that the derivatives of $\pi$ and $\pi'$
\begin{equation}
\partial_\dm \pi^{\da \db} = - L_\dm{}^\dc (a^{-1})_\de{}^\da (a^{-1})_\df{}^\db  f'{}^{\de \df}{}_\dc \qq \partial_\dm' \pi'_{\da \db} = - L'_{\dm \dc} (a'{}^{-1})^\de{}_\da (a'{}^{-1})^\df{}_\db  f_{\de \df}{}^\dc
\end{equation}
can be obtained by deriving the adjoint actions (\ref{Adg}) and (\ref{Adgdual}). Also (\ref{leftrightbrackets}) must be used. }
\begin{equation}\label{FluxesPLFinal}
\begin{aligned}
F_{\dda} &\rightarrow \begin{cases}
F_{\da} &= 0\\
F^{\da} &= (a^{-1})_\dc{}^\da f'{}^{\dc \db}{}_\db\\
\end{cases} \ , \ \ \ F_{\dda \ddb \ddc} \rightarrow
\begin{cases}
F_{\da \db \dc} &= 0\\
F_{\da \db}{}^\dc &= - f_{\da \db}{}^\dc\\
F^{\da \db}{}_\dc &= - {f'}{}^{\da \db}{}_\dc\\
F^{\da \db \dc} &= 0
\end{cases}\\
F'{}_{\dda} &\rightarrow \begin{cases}
F'_{\da} &= (a'{}^{-1})^\dc{}_\da f_{\dc \db}{}^\db\\
F'{}^{\da} &= 0\\
\end{cases} \ , \ \ \ F'{}_{\dda \ddb \ddc} \rightarrow
\begin{cases}
F'{}_{\da \db \dc} &= 0\\
F'{}_{\da \db}{}^\dc &= - f_{\da \db}{}^\dc\\
F'{}^{\da \db}{}_\dc &= - {f'}{}^{\da \db}{}_\dc\\
F'{}^{\da \db \dc} &= 0
\end{cases}
\end{aligned}
\end{equation}
where the structure constants of the bi-algebra (\ref{bialgebrabrackets}) turn out to be the non-vanishing components of the generalized fluxes, as expected. Keeping track of the origin of the fluxes, it can be seen that in the unprimed background the geometric-type fluxes come from $R$-vielbein metric fluxes, while $ \pi $ introduces the non-geometric $ Q $-type flux given by the structure constants $f'$ of the dual algebra. Curiously, in the primed background the $Q$-type fluxes are generated by  $ R' $, and  the geometric ones come from the bi-vector $ \pi' $ (this a generalization of the NATD case where we saw that the dual background consisted of a globally non-geometric space (\ref{Genparalelization}) with a generalized parallelization that rendered the fluxes geometric).

As in the NATD case, we have two different situations. If the groups are unimodular $f_{i j}{}^j = f'_{i j}{}^j = 0$, then the original and dual gaugings fall into the same orbit, both backgrounds are solutions to ungauged DFT, and we are done. If not, the original and dual gaugings (\ref{FluxesPLFinal}) happen to fall into different orbits due to the discrepancy between the vectorial components. Moreover, these gaugings are not constant, as they carry a dependency on the internal coordinates through the adjoint matrices. Interestingly, they still happen to satisfy the consistency constraints (\ref{quadraticconstraints}). The action and equations of motion of DFT depend on the gaugings through the generalized fluxes (\ref{DFTfluxessplit}). Then, the discrepancy between gaugings (\ref{FluxesPLFinal}) can be cured by deforming the original and dual DFT through  shifts in $F_{\cal A}$ intended to annihilate $F_I$  and $F_I'$ respectively. While in the case of NATD these shifts were produced through a gauging procedure (\ref{lambdatilde}), it is unclear to us if similar steps can be taken in this case.
The required deformations again fall into the category of the so-called generalized supergravities \cite{GSE}, as shown in \cite{Thompson PL}, \cite{Sakatani}.  So again, as in the NATD case with non-unimodular gaugings, the local $O(d,d) \times \mathbb{R}^+$ transformation connects solutions of {\it  deformed} DFTs.

Let us point out that as opposed to the dualities considered before, here we start from a non-geometric background (\ref{PLU}) which leads to a non-trivial $\Oma^{(-)}_s$ given by (\ref{Ominusgeneral}) where $u = R$, $\beta = - R^{-t} \pi R^{-1}$ and $b = 0$. This non-geometric behaviour demands that the original background in Gauged DFT is described by a generalized frame in which $e^{(+)} \neq e^{(-)}$.

As mentioned above, Poisson-Lie T-duality is as a generalization of Abelian and non-Abelian T-dualities and so these results must contain both of them as particular cases. Lets see how this works. To do this, we need the explicit infinitesimal expressions for $\pi$ and $\pi'$ which can be obtained using the exponential maps for $g = \exp (Y^i t_i)$ and $g' = \exp (Y'{}_it'{}^i)$ in the definition of the adjoint actions
\begin{equation}\label{formasexplicit}
\begin{aligned}
\pi^{\da \db} =  -{f'}{}^{\da \db}{}_\dc Y^\dc + Y^\dd Y^\de {f'}{}^{\dc [\da}{}_\dd f_{\de \dc}{}^{\db]} - \dots \ , \ \ \  \pi'_{\da \db} =  -{f}{}_{\da \db}{}^\dc Y'_\dc + Y'_\dd Y'_\de {f}{}_{\dc [\da}{}^\dd f'{}^{\de \dc}{}_{\db]} - \dots \ .
\end{aligned}
\end{equation}
For the other objects, namely $L$, $R$ and $a$ and their duals, it will be enough to know that they are trivial for Abelian algebras
\begin{equation}
L_\dm{}^\da = R_\dm{}^\da = \delta_\dm{}^\da \ , \ \ \  a_\da{}^\db = \delta_\da{}^\db \ , \ \ \ L'_{\dm \da} = R'_{\dm \da} = \delta_{\dm \da} \ , \ \ \  a'{}^{\da}{}_\db = \delta^\da{}_\db \ .
\end{equation}

Then, for {\it Abelian} T-duality we have $f = f' = 0$ and so
\begin{equation}
\pi = \pi' = 0 \ , \ \ \  L = R = a = a' = 1  \ , \ \ \ L' = R' = \delta \ .
\end{equation}
Inserting this into (\ref{PLmeasure}) we obtain
\begin{equation}\label{PLsigmasA}
\begin{aligned}
Q'_{\dm \dn} = \delta_{\dm \da} (Q_d^{-1})^{\da \db} \delta_{\db \dn} \ , \ \ \ \Phi' = \Phi - \frac{1}{2}\ln \det {Q_d} \ ,
\end{aligned}
\end{equation}
which are exactly the Abelian transformations (\ref{TdualityPhi}) and (\ref{QprimaAbelian}).

Likewise, for {\it non-Abelian} T-duality (\ref{ENA}) we have $ {f'} = 0 $ but $ f \neq 0 $ so
\begin{equation*}
\pi = 0 \ , \ \ \  {\pi'}_{\da \db} = - f_{\da \db}{}^\dc {Y'}_\dc \ ,
\end{equation*}
Also, since the dual algebra is Abelian $R' = L' = \delta $, $a' = 1$. Inserting this particular case in (\ref{PLmeasure}), we get
\begin{equation}\label{PLsigmasNA}
\begin{aligned}
Q'_{\dm \dn} &= \delta_{\dm \da} R_\dpp{}^\da \left[ \left( \delta^{-1}R^{-1} Q_d + \delta^{-1}f Y' R^t\right){}^{-1}\right]{}^{\dpp}{}_\dn\ ,\\
\Phi' &= \Phi -\frac{1}{2}\ln\left(\text{Det}\left( \delta^{-1} R^{-1}Q_d + \delta^{-1}f Y' R^t\right)\right) + \frac{1}{2}\ln\text{Det}\left(L\right) \ ,
\end{aligned}
\end{equation}
which are the non-Abelian T-dual transformations (\ref{surved_ENA}) and (\ref{NATDPhi}), restricted to the internal sector.

\section{Generalized dualities and higher derivatives}\label{sec_ODDalpha}

In this section we arrive at the main result of this paper: a general formula for first order higher derivative corrections to generalized dualities.

To do so, we implement the procedure described in Section \ref{sec_ODDinDdimensions} to  the next order in $\alpha'$. The path to follow is again the one depicted in Figure \ref{DFTdiagram} in which the corrections will enter trough the double Lorentz transformations. As explained in Section \ref{higherderivatives}, higher derivatives deform the double Lorentz transformations of generalized fields, and consequently of their components. For this reason, the components are not the usual fields in supergravity, but instead are related to them through field redefinitions. In order to distinguish them we use the notation that the components of generalized fields carry an overline  $\bar e ^{(\pm)}$, $\bar B$ and $\bar \Phi$.

Our starting point is again the generalized frame in terms of the supergravity background but now the fields therein are $\alpha'$-corrected, i.e.
\begin{equation}
E_s = \frac 1 {\sqrt{2}}
\begin{pmatrix} - \bar Q^t\,  \bar e^{-t} g^{-1} &  \bar Q \,  \bar e^{-t}\\  \bar e^{-t} g^{-1} &  \bar e^{-t}
\end{pmatrix} \ .
\end{equation}
In order to bring it to a gSS form, we apply a corrected double Lorentz transformation $L_s$. The components of the vielbein $E = L_s(E_s)$ are reached by (\ref{alphaLorentztD})
\begin{equation}\label{LorentzTransfs}
\begin{aligned}
\bar e^{(+)} =& \ \bar e \bar \Oma^{(+)}_s - \cor_s^t G^{-1} e \Oma^{(+)}_s  \\
\bar e^{(-)} =& \ \bar e \bar \Oma^{(-)}_s - \cor_s G^{-1} e \Oma^{(-)}_s \\
\bar G_{GDFT} = L_s(\bar G) =&\ \bar G - (\cor_s + \cor_s^t) \\
\bar B_{GDFT} = L_s(\bar B) =&\ \bar B - (\cor_s - \cor_s^t)  \\
\bar Q_{GDFT} = L_s(\bar Q) =&\ \bar Q - 2\cor_s \\
\bar \Phi_{GDFT} = L_s(\bar \Phi) =& \bar \Phi -\frac{1}{2} \text{Tr}\left(G^{-1}\cor_s\right)\ ,
\end{aligned}
\end{equation}
where $\cor_s = \cor\left(\Oma^{(+)}_s, \Oma^{(-)}_s, \omega^{(\pm)}(e)\right)$ can be read in (\ref{cor}). We are now in the lower-left corner of Figure \ref{DFTdiagram}, so we included a sublabel ``GDFT'' to distinguish the fields from those in the supergravity gauge.

We now move to the right of the diagram in Figure \ref{DFTdiagram} by applying the $O(D,D)$ transformation (\ref{ODDvielbeins}) $E' = T(E) = \Psi E$ together with $d' =T(d) = d$. The results have the same structure as the leading order (\ref{ODDvielbeins}) and (\ref{El4topicante})
\begin{equation}
\bar e^{(\pm)}{}' = T(\bar e^{(\pm)}) \ , \ \ \ \bar Q'_{GDFT} = T(\bar Q_{GDFT}) \ , \  \ \ \bar \Phi'_{GDFT} = T(\bar \Phi_{GDFT})    \ ,
\end{equation}
but now the matrices $\bar \Mma$ and $\bar \Nma$ depend on the overlined fields in Gauged DFT
\begin{equation} \label{MQgdft}
\bar \Mma(\bar Q_{GDFT}) \equiv \Cma \bar Q_{GDFT} + \Dma \ , \ \ \  \bar \Nma (\bar Q_{GDFT}) = -\Cma \bar Q^t_{GDFT} + \Dma \ .
\end{equation}
The matrices $\Ama$, $\Bma$, $\Cma$ and $\Dma$ in the $O(D,D)$ element in (\ref{ODDg}) receive no corrections, and so remain unbarred. The same happens with the generalized dilaton shift $\alpha$ (\ref{gendilatonshift}). We now work a little on (\ref{MQgdft})
\begin{equation}\label{mmacor}
\begin{aligned}
\bar \Mma(\bar Q_{GDFT}) &= \bar \Mma (\bar Q) - 2 \Cma \cor_s \ \ \Rightarrow \ \ \bar \Mma(\bar Q_{GDFT})^{-t} = \bar \Mma^{-t} + 2 \Mma^{-t}\cor_s^t\Cma^t \Mma^{-t}\\
\bar \Nma(\bar Q_{GDFT}) &= \bar \Nma (\bar Q) + 2 \Cma \cor_s^t \ \ \ \Rightarrow \ \ \bar \Nma(\bar Q_{GDFT})^{-t} = \bar \Nma^{-t} - 2 \Nma^{-t}\cor_s\Cma^t \Nma^{-t} \ ,
\end{aligned}
\end{equation}
where we used the identity $(A + \epsilon)^{-1} \eqsim A^{-1} - A^{-1} \epsilon A^{-1}$ for small perturbations, and truncated the result to first order in $ \alpha' $. Then, the matrices $\bar \Mma$ and $\bar \Nma$ appearing now depend on the background $\bar Q$ in the supergravity double Lorentz gauge. Introducing these expressions into $\bar e^{(\pm)}{}'$ together with the explicit form of $\bar e^{(\pm)}$ in terms of $\bar e$ in (\ref{LorentzTransfs}) we get
\begin{equation}
\begin{aligned}
\bar e^{(+)}{}' &= \bar \Mma^{-t}\bar e \bar \Oma^{(+)}_s - \Mma^{-t}\cor_s^t\left(1 - 2 \Cma^t \Mma^{-t} G \right) G^{-1} e \Oma^{(+)}_s\\
\bar e^{(-)}{}' &= \bar \Nma^{-t}\bar e \bar \Oma^{(-)}_s - \Nma^{-t}\cor_s\left(1 + 2 \Cma^t \Nma^{-t} G \right) G^{-1} e \Oma^{(-)}_s  \ .
\end{aligned}
\end{equation}

These expressions can be improved by using the following $ O(D,D) $ identities
\begin{equation}\label{MNCidentities}
\begin{aligned}
\Cma^t = -\Mma^{-1} \Cma \Nma^t \ , \ \ \   \Mma = \Nma + 2 \Cma G \ , \ \ \ \Nma^{-1}\Mma = 1 - 2 \Cma^t \Mma^{-t} G \ , \ \ \ \Mma^{-1}\Nma = 1 + 2 \Cma^t \Nma^{-t} G \ .
\end{aligned}
\end{equation}
Using these identities we arrive at
\begin{equation}
\begin{aligned}
\bar e^{(+)}{}' &= \bar \Mma^{-t}\bar e \bar \Oma^{(+)}_s - \Mma^{-t}\cor_s^t \Nma^{-1} \Mma G^{-1} e \Oma^{(+)}_s\\
\bar e^{(-)}{}' &= \bar \Nma^{-t}\bar e \bar \Oma^{(-)}_s - \Nma^{-t}\cor_s\Mma^{-1} \Nma G^{-1} e \Oma^{(-)}_s\\
\bar G'_{GDFT} &= \bar \Mma^{-t} \bar G \bar \Mma^{-1} - \Mma^{-t} \cor_s^t \Nma^{-1} - \Nma^{-t} \cor_s \Mma^{-1}\\
\bar B'_{GDFT} &= \bar \Mma^{-t} \bar B^* \bar \Mma^{-1} + \Mma^{-t} \cor_s^t \Nma^{-1} - \Nma^{-t} \cor_s \Mma^{-1}\\
\bar Q'_{GDFT} &= \bar \Mma^{-t} \bar Q^* \bar \Mma^{-1} - 2 \Nma^{-t} \cor_s \Mma^{-1}\\
\bar \Phi'_{GDFT} &= \bar \Phi - \frac{1}{2} \ln \det {\bar \Mma} + \alpha -\frac{1}{2} \text{Tr}\left(\Nma G^{-1}\cor_s \Mma^{-1}\right)\ .
\end{aligned}
\end{equation}
These equations express the fields in the lower-right corner of Figure \ref{DFTdiagram} in terms of those in the upper-left corner.

Next we implement the last arrow in Figure \ref{DFTdiagram} with a corrected double Lorentz $L_s'$ to arrive at the dual supergravity gauge $E_s'$
\begin{equation}
\begin{aligned}
\bar e' &= \bar \Mma^{-t}\bar e \bar \Oma^{(+)}_s \Oma_s^{(+)}{}' - \left[\Mma^{-t}\cor_s^t \Nma^{-1} + \cor'_s{}^{t}\right] G'{}^{-1} \Mma^{-t} e \Oma^{(+)}_s\Oma_s^{(+)}{}'\\
&= \bar \Nma^{-t}\bar e \bar \Oma^{(-)}_s \Oma_s^{(-)}{}' - \left[\Nma^{-t}\cor_s \Mma^{-1} + \cor'_s\right] G'{}^{-1} \Nma^{-t} e \Oma^{(-)}_s\Oma_s^{(-)}{}'\\
\bar G' &= \bar \Mma^{-t} \bar G \bar \Mma^{-1} - \left[\Mma^{-t}\cor_s^t \Nma^{-1} + \cor'_s{}^{t}\right] - \left[\Nma^{-t}\cor_s \Mma^{-1} + \cor'_s\right]\\
\bar B' &= \bar \Mma^{-t} \bar B^* \bar \Mma^{-1}  + \left[\Mma^{-t}\cor_s^t \Nma^{-1} + \cor'_s{}^{t}\right] - \left[\Nma^{-t}\cor_s \Mma^{-1} + \cor'_s\right]\\
\bar Q' &= \bar \Mma^{-t} \bar Q^* \bar \Mma^{-1}  -2 \left[\Nma^{-t}\cor_s \Mma^{-1} + \cor'_s\right]\\
\bar \Phi' &= \bar \Phi - \frac{1}{2} \ln \det {\bar \Mma} + \alpha -\frac{1}{2} \text{Tr}\left(G'{}^{-1}\left[\Nma^{-t}\cor_s \Mma^{-1} + \cor'_s\right]\right)\ ,
\end{aligned}
\end{equation}
where $\cor_s' = \cor\left(\Oma^{(+)}_s{}', \Oma^{(-)}_s{}', \omega^{(\pm)}(e^{(\pm)}{}')\right)$.

Finally, we choose the same gauge we took for the leading order
\begin{equation}
    \bar e^{(+)} = e \ , \ \ \ \bar e^{(+)}{}' = \bar e' \ ,
\end{equation}
so that
\begin{equation}
    \bar \Oma_s^{(+)} = \bar \Oma_s^{(+)}{}' = 1 \ ,
\end{equation}
and
\begin{equation}
\bar \Oma_s^{(-)} \bar \Oma_s^{(-)}{}' = \bar e^{-1} \bar \Nma^t \bar \Mma^{-t} \bar e + e^{-1} \Nma^t \left[\left(\Nma^{-t}\cor_s \Mma^{-1} + \cor'_s\right)  - \left(\Nma^{-t}\cor_s \Mma^{-1} + \cor'_s\right){}^t\right] G'{}^{-1} \Mma^{-t} e    \ ,
\end{equation}
which can be shown to be a Lorentz element as a consequence of the antisymmetric appearance of $\left(\Nma^{-t}\cor_s \Mma^{-1} + \cor'_s\right)$.  In this gauge, where $    \bar \Oma_s^{(+)} = \bar \Oma_s^{(+)}{}' = 1$, we have that $\cor^{(+)} = 0$ both in $\cor_s$ and $\cor_s'$ as it should be clear from (\ref{cor}). This leaves $\cor^{(-)}$ as the only contribution and so the dependency on the parameter $b$ is completely removed. Starting with a different diagonal Lorentz gauge in the supergravity gauge would lead to other dependence on the parameters $a$ and $b$. In particular if we started in the gauge in which $e = e^{(-)}$, that would induce corrections with only the parameter $b$.

We have then finally arrived at the first order in $\alpha'$ generalized T-duality transformations in the DFT scheme:
\begin{boxquation}
\begin{equation}\label{alphageneralizedduality}
\begin{aligned}
\bar e' &= \bar\Mma^{-t} \bar e - \left[\Mma^{-t}\cor_s^t \Nma^{-1} + \cor'_s{}^{t}\right] \Mma G^{-1} e\\
\bar G' &= \bar \Mma^{-t} \bar G \bar \Mma^{-1} - \left[\Mma^{-t}\cor_s^t \Nma^{-1} + \cor^{t}{}'_s\right] - \left[\Nma^{-t}\cor_s \Mma^{-1} + \cor'_s\right]\\
\bar B' &= \bar \Mma^{-t} \bar B^* \bar \Mma^{-1} + \left[\Mma^{-t}\cor_s^t \Nma^{-1} + \cor'_s{}^{t}\right] - \left[\Nma^{-t}\cor_s \Mma^{-1} + \cor'_s\right] \\
\bar Q' &= \bar \Mma^{-t} \bar Q^* \bar \Mma^{-1} - 2 \left[\Nma^{-t}\cor_s \Mma^{-1} + \cor'_s\right]\ , \ \
\bar Q^* = \bar G + \bar B^* \ , \ \  \bar B^* = \bar B + \Dma^t  \Bma + \bar Q^t\Cma^t\Ama \bar Q + \bar Q^t\Cma^t\Bma - \Bma^t \Cma \bar Q\\
\bar \Phi' &= \bar \Phi - \frac{1}{2} \ln \det {\bar \Mma} + \alpha -\frac{1}{2} \text{Tr}\left(G'{}^{-1}\left[\Nma^{-t}\cor_s \Mma^{-1} + \cor'_s\right]\right)\ ,
\end{aligned}
\end{equation}
\end{boxquation}

where $\cor_s = \cor\left(1, \Oma^{(-)}_s, \omega^{(-)}(e)\right)$ and $\cor_s' = \cor\left(1, \Oma^{(-)}_s{}', \omega^{(-)}(e^{(-)}{}')\right)$ with $\Oma_s^{(-)}$ and $\Oma_s^{(-)}{}'$ defined in (\ref{Ominusgeneral}). The dependency on $\omega^{(-)}(e^{(-)}{}')$ can be improved using $ e^{(-)}{}'= \Nma^{-t} e^{(-)}  = \Nma^{-t} e \Oma_s^{(-)}$ and the Lorentz and $O(D,D)$ transformations of $\omega^{(-)}$ in (\ref{Lomega}) and (\ref{Dualityomega}) respectively
\begin{equation}
\omega^{(-)}(e^{(-)}{}') = \Mma^{-t} \left(\Oma_s^{(-) -1} \omega^{(-)}(e) \Oma_s^{(-)}  + \Oma_s^{(-)-1} \partial \Oma_s^{(-)}\right) \ .
\end{equation}

These are the first order corrections to the equations (\ref{Los3picantes}) and (\ref{El4topicante}). They capture any generalized duality, encoded here in generic local $O(D,D) \times \mathbb{R}^+$ transformations, for any choice of the parameters $a$ and $b$ that control the first-order corrections in the deformed DFT. These expressions are valid in the DFT scheme, namely for the components of the duality covariant fields after the gauge fixing. These are {\it not} the fields that appear in supergravity, but are related to them through field redefinitions, as we discuss in the following section. Note that the right hand side in equation (\ref{alphageneralizedduality}) contains the original background in the DFT scheme $\bar e$, $\bar B$, $\bar \Phi$. All the other elements that appear ($\mathbb{A}$, $\mathbb{B}$, $\mathbb{C}$, $\mathbb{D}$, $\alpha$, $\Oma^{(-)}_s$, $\Oma^{(-)}_s{}'$) can be read from the generalized duality to the lowest order. So knowing how the duality works to lowest order, and having a corrected supergravity solution permits to compute its dual from this expression.

In the following section we will need the generic Lorentz transformation of $\bar e$. From the DFT point of view the double Lorentz transformations acts differently on $\bar e^{(+)}$ and $\bar e^{(-)}$, so in order for this transformation to keep us in the supergravity gauge we need $L(\bar e^{(+)}) = L(\bar e^{(-)}) = L(\bar e)$ in (\ref{alphaLorentztD}). This forces a relation between $\bar \Oma^{(+)}$ and $\bar \Oma^{(-)}$
\begin{equation}
\bar \Oma^{(-)} = \bar \Oma^{(+)} + e^{-1}\left(\cor  - \cor^t \right)e^{-t} g^{-1} \Oma^{(+)}  \ ,
\end{equation}
that is solved as follows
\begin{equation}
\begin{aligned}
&\bar \Oma^{(-)} = (1 + \gamma A g^{-1}) \, \bar  \Oma \ ,  \ \ \ \ \ \ \ \ \ \ A = e^{-1}\left(\cor  - \cor^t \right)e^{-t}\\
&\bar \Oma^{(+)} = (1 + (\gamma - 1) A g^{-1}) \, \bar \Oma \ . \\
\end{aligned}
\end{equation}
It is clear that these are elements of the Lorentz group for any value of the parameter $\gamma$ because the matrix $A$ is antisymmetric. Note that while to lowest order the two elements are forced to coincide (this is the usual case in which the double Lorentz symmetry breaks to its diagonal subgroup), higher orders make the two transformations differ. Since $\gamma$ can be chosen at will, we set its value to $\gamma = 1$. This implies the following Lorentz transformations for the gauge fixed fields
\begin{equation}\label{Lorentztransformationofbarredfields}
\begin{aligned}
&L(\bar e) = \bar e \, \bar \Oma - \cor^t G^{-1} e\,  \Oma  \\
&L(\bar G) = \bar G - (\cor + \cor^t) \\
&L(\bar B) = \bar B - (\cor - \cor^t)  \\
&L(\bar Q) = \bar Q - 2 \cor \\
&L(\bar \Phi) = \bar \Phi -\frac{1}{2} G^{\Dm \Dn} \cor_{\Dm \Dn}\ ,
\end{aligned}
\end{equation}
where $\cor = \cor(\Oma, \Oma, \omega^{(\pm)}(e))$ is given in (\ref{cor}). Other choices of $\gamma$ simply ammount to redefinitions of $\bar \Oma$.
These expressions are important, because as opposed to this DFT supergravity scheme, in all other supergravity schemes the metric and dilaton  are Lorentz invariant, and then field redefinitions will be required to remove this anomalous transformation.

We explained at the end of Section \ref{higherderivatives} why, even at higher orders, the local $O(D,D)\times \mathbb{R}^+$ transformations map solutions into solutions of DFT. In the context of Gauged DFT this is realized rather trivially: the transformation keeps the gaugings into the same orbit and then works as a symmetry of the Gauged DFT. Even if the orbits are different one can make sense of the transformation as a solution generating technique between deformed theories, as we discussed for instance when gaugings are non-unimodular. It is then natural to ask why this extreme simplicity is no longer reflected in the results of this section. The reason is that the gauge choice necessary to make contact with supergravity (in the DFT scheme) requires double Lorentz transformations which are deformed by higher derivatives.

\subsection{Supergravity schemes}

The overline on fields in the previous section indicates that they are components of the generalized fields in DFT, and so we call this set of fields the DFT scheme. In this scheme the frame field receives a first order  Lorentz transformation inherited from the generalized Green-Schwarz transformation (\ref{Lorentztransformationofbarredfields}), and so it is {\it not} the standard frame field in supergravity. However, it is related to it through a first order Lorentz non-covariant field redefinition. The same is true for the dilaton and two-form (although in some cases the Lorentz transformation of the two-form cannot be redefined away). So the fields in the DFT scheme (with an overline) and the fields in supergravity (without an overline) are related by
\begin{equation}
\begin{aligned} \label{fieldexpansion}
\bar e &= e +  \Delta e \ , \ \ \  \bar B &= B +  \Delta B \ , \ \ \  \bar \Phi &= \Phi + \Delta \Phi \ .
\end{aligned}
\end{equation}
 The correction $\Delta$  depends on the supergravity scheme to be considered, and is defined up to covariant Lorentz redefinitions. The non-covariant part is fixed by
\begin{equation} \label{LorentzCovariant}
L(e) = e \, \Oma \ , \ \ \ \ L(G) = G \ , \ \ \ \ L(\Phi) = \Phi    \ .
\end{equation}
The only case in which the two-form can be taken to be a Lorentz invariant field $L(B) = B$ is when $a = b$, which corresponds to the bosonic string \cite{MarquesNunez}. Otherwise it carries a Green-Schwarz transformation. Different supergravity schemes \cite{schemes}-\cite{BdR} correspond to different choices of ($\Delta e$, $\Delta B$, $\Delta \Phi$) related by Lorentz covariant field redefinitions.

Applying a generalized duality to $\bar e'$ leads to
\begin{equation}
\bar e' = e' + (\Delta e)'  \ \ \ \Rightarrow \ \ \ e' = \bar e' - (\Delta e)' \ ,
\end{equation}
where $(\Delta e)' \equiv \Delta e (e')$. We know from (\ref{alphageneralizedduality}) what $\bar e'$ is in terms of $\bar e$, and from (\ref{fieldexpansion}) what $\bar e$ is in terms of $e$, so we can readily compute $e'$ expanding $\bar \Mma^{-t} = \Mma^{-t} - \Mma^{-t} (\Delta Q)^t \Cma^t \Mma^{-t}  $ in the same way we did in (\ref{mmacor}), we then have
\begin{equation}
\begin{aligned}
e' = \Mma^{-t} e + \Mma^{-t} \Delta e - \Mma^{-t} (\Delta Q)^{t}\Cma^t \Mma^{-t} e - \left[\Mma^{-t}\cor_s^t \Nma^{-1} + \cor'_s{}^{t}\right] \Mma G^{-1} e - (\Delta e)' \ .
\end{aligned}
\end{equation}

For the metric the above results imply
\begin{equation}\label{Gres}
\begin{aligned}
G' &= G^{(0)} + G^{(1)}\\
G^{(0)} &= \Mma^{-t} G \Mma^{-1}\\
G^{(1)} &= \frac{1}{2}\Nma^{-t}\Delta Q \Mma^{-1} - \left[\Nma^{-t}\cor_s \Mma^{-1} + \cor'_s\right] - \frac{1}{2}(\Delta G)'  + \text{Transpose}\ ,
\end{aligned}
\end{equation}
where we used the identities (\ref{MNCidentities}). For the two-form we follow the same procedure and after introducing $ B^* $ as in (\ref{Bstar}) and using exhaustively the $ O(D,D) $ identities, we can get a similar result as for the metric
\begin{equation} \label{Bres}
\begin{aligned}
B' &= B^{(0)} + B^{(1)}\\
B^{(0)} &= \Mma^{-t} B^* \Mma^{-1}\\
B^{(1)} &= \frac{1}{2} \Nma^{-t}\Delta Q \Mma^{-1} - \left[\Nma^{-t}\cor_s \Mma^{-1} + \cor'_s\right] - \frac{1}{2}(\Delta B)' - \text{Transpose}\ .
\end{aligned}
\end{equation}

Both results (\ref{Gres}) and (\ref{Bres}) can then be merged into a single expression in terms of $ Q' = Q^{(0)} + Q^{(1)}$.  The final result for first order corrections to generalized dualities is given by:
\begin{boxquation}
	\begin{equation} \label{ResultadoFinal}
	\begin{aligned}
	e^{(0)} &= \Mma^{-t} e \ , \ \ \ \ \ \ \ \ \ \ \ \ \ \ e^{(1)} = \Mma^{-t} \Delta e - \Mma^{-t} (\Delta Q)^{t}\Cma^t \Mma^{-t} e - \left[\Mma^{-t}\cor_s^t \Nma^{-1} + \cor^{t}{}'_s\right] \Mma G^{-1} e - (\Delta e)' \\
	Q^{(0)} &= \Mma^{-t} Q^* \Mma^{-1} \ , \ \ \ \  Q^{(1)} = \Nma^{-t}\left(\Delta Q - 2 \cor_s\right) \Mma^{-1}  - (\Delta Q)' -2 \cor'_s\\
	\Phi^{(0)}&= \Phi - \frac{1}{2} \ln \text{Det}(\Mma) + \alpha \\
	\Phi^{(1)} &= -\frac{1}{2} \text{Tr}\left(\Mma^{-1}\Cma \Delta Q + G^{-1'}\left[\Nma^{-t}\cor_s \Mma^{-1} + \cor'_s\right]\right) + \Delta \Phi - (\Delta \Phi)'\ .
	\end{aligned}
	\end{equation}
\end{boxquation}
We have then extended the result of the previous subsection to be applicable to generic schemes related by field redefinitions from the DFT scheme. This reduces to (\ref{alphageneralizedduality}) when  $\Delta e = \Delta B = \Delta \Phi = 0$.

The fields without an overline must transform covariantly under Lorentz transformations (\ref{LorentzCovariant}). We can then separate $\Delta$ into a non-covariant part, and a covariant part. The former is unambiguously defined, and the scheme in which $\Delta$ contains only the non-covariant part was named the Bergshoeff-de Roo (BdR) scheme in \cite{MarquesNunez}, after \cite{BdR}
\begin{equation}\label{BdRscheme}
\Delta G^{(\text{BdR})}_{\Dm \Dn} = -\frac{1}{4}\left(a\omega^{(-) 2}_{\Dm \Dn} + b \omega^{(+) 2}_{\Dm \Dn}\right)\ , \ \ \
\Delta \Phi^{(\text{BdR})} = \frac{1}{4} G^{\Dm \Dn} \Delta G_{\Dm \Dn}\ , \ \ \
\Delta B^{(\text{BdR})}_{\Dm \Dn} = 0 \ ,
\end{equation}
with $ \omega^{(\pm)2}_{\Dm \Dn} = \omega^{(\pm)}_{\Dm \DFa}{}^\DFb \omega^{(\pm)}_{\Dn \DFb}{}^\DFa $. As explained, it is not always possible to make the two-form Lorentz invariant, and interestingly in the BdR scheme the two-form coincides with the two-form in the DFT scheme.

\subsection{Examples}

\subsubsection{Abelian T-duality}

The $ \alpha'$-corrected T-duality transformations must contain the corrections to Abelian T-duality as a particular case. In order to check this statement, we consider the decomposition of $O(D,D)$ into $ GL(D) $ transformations, $ B$-shifts and factorized T-dualities. In the three cases the matrices $ \Ama,\Bma,\Cma,\Dma $ are constant, and we take the generalized dilaton shifts to vanish $\alpha = 0$. Moreover, we remember that in this case $\Oma^{(-)}_s = 1$.

For $ GL(D) $ transformations, we have $ \Bma=\Cma = 0 $ and $  \Dma = \Ama^{-t} $ so
\begin{equation}
\Mma = \Nma = \Ama^{-t} \qq \Rightarrow \qq  \Oma^{(-)}_s{}'  = 1 \ .
\end{equation}
For $ B$-shifts $ \Ama=\Dma = 1 $, $ \Bma = \text{constant}$ and $ \Cma = 0 $ so
\begin{equation}
\Mma = \Nma = 1 \qq \Rightarrow \qq \Oma^{(-)}_s{}'  = 1 \ .
\end{equation}
For a single factorized T-duality in a particular direction  $ x $, we need the reduced form of the matrices (\ref{embedding}). In this case $ a=d=0$ and $ b=c=1 $ with
\begin{equation}
\mma = - \nma = G_d  \qq \Rightarrow \qq \oma^{(-)}_s{}' = -1 \ .
\end{equation}
Then, in the three cases $\Oma^{(-)}_s$ and $\Oma^{(-)}_s{}'$ are constant, and consequently $\cor_s = \cor'_s = 0 $. This reduces the general formulas (\ref{ResultadoFinal}) to
\begin{equation} \label{transfAbelianDFTs}
\begin{aligned}
Q^{(0)} &= \Mma^{-t} Q \Mma^{-1} \ , \ \ \ \ \ \ \ \ \ \ \  Q^{(1)} = \Nma^{-t} \Delta Q \Mma^{-1} - (\Delta Q)'\\
\Phi^{(0)}&= \Phi - \frac{1}{2} \ln \text{Det}(\Mma) \ , \ \ \  \Phi^{(1)} = -\frac{1}{2}\text{Tr}\left(\Mma^{-1}\Cma\Delta Q \right) + \Delta \Phi - (\Delta \Phi)' \\
\omega^{(+)}{}' &= \Nma^{-t} \omega^{(+)} \ , \ \ \ \  \ \ \ \ \ \ \ \ \ \ \
\omega^{(-)}{}' = \Oma^{(-)}_s{}^{-1}{}' \Mma^{-t} \omega^{(-)} \Oma^{(-)}_s{}' \ ,
\end{aligned}
\end{equation}
where the transformations $\omega^{(\pm)}{}'$ are obtained from (\ref{Omegaprime}) with $\Oma^{(-)}_s = 1 $ and $\Oma^{(-)}_s{}' = \text{Constant}$.

To move forward, we need to specify a particular scheme. In this case we will consider the Bergshoeff-de Roo scheme with the field redefinitions given in (\ref{BdRscheme}). The transformed version of those field redefinitions can be easily obtained from $\omega^{(\pm)}{}'$
\begin{equation}
\omega^{(+)2} = \Nma^{-t} \omega^{(+)2} \Nma^{-1} \ , \ \ \ \omega^{(-)2} = \Mma^{-t} \omega^{(-)2} \Mma^{-1} \ .
\end{equation}
Introducing these objects together with (\ref{BdRscheme}) in our general formula, and after some work using $O(D,D)$ identities, we arrive at
\begin{equation}
Q^{(1)} =\frac{1}{2}\Nma^{-t}\left[ a G \Nma^{-1} \Cma \omega^{(-)2}  - b \omega^{(+)2} \Cma^t \Mma^{-t} G \right]\Mma^{-1} \ , \ \ \Phi^{(1)} = \frac{1}{8}(\Mma^{-1})^\Dm{}_\Dn \Cma^{\Dn \Dr}\left( a\omega^{(-)2}_{\Dr \Dm} + b\omega^{(+)2}_{\Dr \Dm} \right)  \ .
\end{equation}
For $ GL(D) $ transformations and $ B$-shifts we have $ \Cma = 0 $ and so $ Q^{(1)} = \Phi^{(1)} = 0 $, so interestingly $ GL(D) $ and $ B$-shifts receive no corrections.
For  factorized T-dualities instead we expect higher derivative corrections. Consider the heterotic string in particular, for which $ a=0 $ and $ b=-1 $
\begin{equation*}
\begin{aligned}
Q^{(1)} = \frac{1}{2}\Nma^{-t}\omega^{(+)2} \Cma^t \Mma^{-t} G\Mma^{-1} \ , \ \ \  \Phi^{(1)} = -\frac{1}{8}(\Mma^{-1})^\Dm{}_\Dn \Cma^{\Dn \Dr} \omega^{(+)2}_{\Dr \Dm} \ , \ \ \  \omega^{(+)}{}' = \Nma^{-t} \omega^{(+)} \ .
\end{aligned}
\end{equation*}
We consider the simple case of a single internal isometric direction, and then perform a splitting as we did in Section \ref{sec_reduction} for the zeroth order. In this case the results are
	\begin{equation}
	\begin{aligned}
	\bar Q_{\nii \nj}' &= \bar Q_{\nii \nj} - \frac{\bar Q_{\nii x} \bar Q_{x \nj}}{\bar Q_{x x}} \ , \ \ \  \bar Q_{x x}' = \frac{1}{\bar Q_{x x}}	\ , \ \ \  \bar Q_{\nii x}' = \frac{\bar Q_{\nii x}}{\bar Q_{x x}} \ , \ \ \  \bar Q_{x \nii} = - \frac{\bar Q_{x \nii}}{\bar Q_{x x}}\\
	\bar e'_{\nii}{}^{\alpha} &= \bar e_{\nii}{}^\alpha - \frac{\bar Q_{x \nii}}{\bar Q_{x x}} \bar e_{x}{}^\alpha \ , \ \ \ \bar e'_{x}{}^{\alpha} = \frac{\bar e_{x}{}^{\alpha}}{\bar Q_{x x}}\\
	 \bar \omega^{(+)}{}'_{\nii \alpha}{}^{\beta} &=   \bar \omega^{(+)}_{\nii \alpha}{}^{\beta} - \frac{\bar Q_{\nii x}}{\bar Q_{x x}} \bar \omega^{(+)}_{x \alpha}{}^{\beta} \ , \ \ \  \bar  \omega^{(+)}{}'_{x \alpha}{}^{\beta} = - \frac{ \bar \omega^{(+)}_{x \alpha}{}^{\beta}}{\bar Q_{x x}}\\
	G_{\nii \nj}' &= G_{\nii \nj} - \frac{1}{G_{x x}} (G_{\nii x} G_{\nj x} + b_{\nii x} B_{\nj x}) + \frac{1}{G^2_{x x}}\left(G_{x x} \Omega_{x (\nii} B_{\nj) x}  - \Omega_{x x} G_{x (\nii}B_{\nj) x} - \Omega_{x x} B_{\nii x}B_{\nj x} \right)\\
	G_{\nii x}' &= \frac{B_{\nii x}}{G_{x x}} + \frac{1}{2 G^2_{x x}}\left( G_{x x} \Omega_{\nii x}  - \Omega_{x x} G_{\nii x} - 2 \Omega_{x x} B_{\nii x}\right)\\
	G_{x x}' &= \frac{1}{G_{x x}} -\frac{\Omega_{x x}}{G^2_{x x}}\\
	B_{\nii \nj}' &= B_{\nii \nj}  - \frac{1}{G_{x x}} (G_{\nii x} B_{x \nj} + B_{\nii x} G_{x \nj}) + \frac{1}{G^2_{x x}}\left(G_{x x} \Omega_{x [\nii} B_{\nj] x}  - \Omega_{x x} G_{x [\nii}B_{\nj] x}\right) \\
	B_{\nii x}' &= \frac{G_{\nii x}}{G_{x x}} + \frac{1}{2 G^2_{x x}}\left( G_{x x} \Omega_{\nii x}  - \Omega_{x x} G_{\nii x}\right)\\
	\Phi' &= \Phi - \frac{1}{2}\ln G_{x x} - \frac{1}{4}\frac{\Omega_{x x}}{G_{x x}} \ ,
	\end{aligned}
	\end{equation}
where we defined $ \Omega \equiv \frac{1}{2} \omega^{(+)2} $. These results are the same as the ones obtained in \cite{Ortin} for the heterotic string after identifying $ B_{\nii x}^{(\text{here})} = - B_{\nii x}^{(\text{there})} $ and setting the $ \alpha $ parameter in that paper to $ \frac{1}{2} $ (see eqs. (39,42,70,74,75,76)).

\subsubsection{Yang-Baxter}

We now move to a different generalized duality for backgrounds with non-Abelian isometries. In \cite{BW1} it was shown that after applying unimodular homogeneous YB transformations over bosonic string solutions at order $\alpha'$, the resulting background could be corrected to satisfy the equations of motion.  This was done for backgrounds with vanishing NSNS fluxes and up to second order in the deformation parameter $\eta$. Soon after, in \cite{BW2} it was realized that the same result could be obtained by considering these particular generalized dualities in the context of DFT to order $\alpha'$. In this case the deformed background was obtained at all orders in $\eta$ and the original background was allowed to have NSNS-fluxes. As expected, the result reduced to the previous one after setting the particular conditions of \cite{BW1}.

Our general formula for higher-derivative corrections to generalized dualities includes this scenario as a particular case. We will show here that the results of \cite{BW2} are recovered, a task that will turn out easy because we are using a notation similar to the one used there. To see this, we first  notice that (\ref{YB}) can be trivially extended to $ D$-dimensions by
\begin{equation}
\Theta^{\dm \dn} \rightarrow \Theta^{\Dm \Dn} = k_\da{}^\Dm \text{R}^{\da \db} k_\db{}^\Dn \ ,
\end{equation}
where $ k_\da{}^\Dm $ are extended by introducing the identity map on the external directions. The same can be done for the Maurer-Cartan form and so the expression (\ref{YB}) can be brought to
\begin{equation}
Q' = Q\left(\eta \Theta Q + 1\right) \ , \ \ \  \Phi' = \Phi - \frac{1}{2}\ln\left(\text{Det}(\eta \Theta Q + 1)\right) \ .
\end{equation}
Then, we specify our results (\ref{ResultadoFinal}) to the bosonic case $a=b=-1$ and consider the  scheme used in \cite{BW1}
\begin{equation} \label{HTscheme}
\Delta Q_{\Dm \Dn} = \frac{1}{2} \omega^{(+)}_{\Dm \DFa}{}^{\DFb} \omega^{(-)}_{\Dn \DFb}{}^{\DFa} \ , \ \ \  \Delta \Phi =  -\frac{1}{48} H^2 + \frac{1}{4} G^{\Dm \Dn} \Delta G_{\Dm \Dn} \ ,
\end{equation}
where $H^2 \equiv H_{\Dm \Dn \Dr} H^{\Dm \Dn \Dr}$. Remembering that we come from a geometric background in which $e^{(+)} = e^{(-)}$ and consequently $\Oma_s^{(-)} = 1$ and $\cor_s^{(-)} = 0$,  the corrected unimodular homogeneous YB transformation is then obtained
\begin{eqnarray}
Q^{(1)} &=& -\frac{1}{2} \omega^{(-)}{}'{}_{\Dn \DFa}{}^\DFb\left(\omega^{(+)}{}'{}_{\Dm \DFb}{}^\DFa - \Oma_s^{(-)-1}{}'{}_{\DFb}{}^\DFc \partial_{\Dm}  \Oma_s^{(-)}{}'{}_{\DFc}{}^\DFa \right)
+ \frac{1}{4} \partial_\Dm  \Oma_s^{(-)-1}{}'{}_\DFa{}^\DFb \partial_\Dn \Oma_s^{(-)}{}'{}_{\DFb}{}^\DFa
+ \frac{1}{2}\cor^{\text{WZW}}(\Oma_s^{(-)}{}'){}_{\Dm \Dn}\nonumber \\
&& + \frac{1}{2}\left[(-Q^t \Theta + 1)^{-1}\right]{}_\Dn{}^\Dr \omega^{(-)}{}_{\Dr \DFa}{}^\DFb \omega^{(+)}{}_{\Ds \DFb}{}^\DFa\left[(-\Theta Q^t + 1)^{-1}\right]{}^\Ds{}_\Dm \\
\omega^{(+)}{}' &=& \Nma^t \omega^{(+)} \ , \ \ \
\omega^{(-)}{}' = \Oma_s^{(-)}{}^{-1}{}' \Mma^{-t} \omega^{(-)} \Oma_s^{(-)}{}' + \Oma_s^{(-)}{}^{-1}{}'\partial \Oma_s^{(-)}{}' \ , \ \ \
\Oma_s^{(-)}{}' = e^{-1} \Nma^t \Mma^{-t} e \ , \nonumber
\end{eqnarray}
where the transformation for $\omega^{(\pm)}$ follows from (\ref{Omegaprime}) with $\Oma_s^{(-)}=1$.

From here, we can see after the change $\Theta \rightarrow - \Theta$, $\Oma_s^{(-)}{}' \rightarrow  \Oma_s^{(-)}{}^{-1}{}'$ these are the same results obtained in eqs. (3.6), (4.15), (4.25) and (4.25) of \cite{BW2}. Finally, for the dilaton field instead of using our general formula, the more straightforward way to match results is noticing that in the scheme (\ref{HTscheme}) one has
\begin{equation}
e^{-2 d} = e^{-2 \bar \Phi} \sqrt{\bar G} = e^{-2 \Phi} \sqrt{G}\left(1 + \frac{1}{24} H^2\right) \ ,
\end{equation}
so using that for YB the generalized dilaton shift vanishes (\ref{YBQ2}) $d' = d$ we get the transformation for the dilaton
\begin{equation}
\Phi' = \Phi - \frac{1}{2} \ln \Mma + \frac{1}{4} G^{(0) \Dm \Dn} G^{(1)}_{\Dm \Dn} + \frac{1}{48}\left( H'{}^{2} - H^{2}\right)\ ,
\end{equation}
which is exactly the expression given there in eq. (4.16).

\section{Outlook}

A number of questions arise:

\begin{itemize}
    \item {\bf Explicit solutions.} It would be interesting to apply our result to specific examples. Higher-derivative corrections to Abelian T-duality have been applied in different contexts, such as corrections to entropy and black-hole solutions \cite{Edelstein}, \cite{OrtinBH} and cosmological backgrounds \cite{HZallorders}. Higher-derivative corrections to Yang-Baxter deformations were recently considered in \cite{BW2}. There have also been some analysis on higher-derivative corrections to non-Abelian and PL dualities \cite{PLandalpha}.

    \item {\bf Classification of generalized dualities.} An interesting observation is that the framework of Gauged DFT allows to envision further extensions of generalized dualities, beyond those discussed here. In particular it might offer a classification through classifications of duality orbits  on the one hand, and on the other through the characterization of the degeneracy in the space of duality twists that fall into the same orbit. Steps in this direction were given in \cite{DualityOrbits} and \cite{Inverso}. Also the formalism of DFT$_{\rm WZW}$ \cite{DFTWZW} can be useful in this respect because the frame algebra is simpler, and gives a prescription to compute generalized twists in Gauged DFT.
    There are a priori no obstructions in finding examples of generalized dualities in Gauged DFT that go beyond PL T-plurality. An interesting case of study is the so called $\mathcal{E}$-models \cite{Emodels} recently discussed in the context of DFT in \cite{DoubleAspects}.

    \item {\bf Extensions to higher orders.} The whole construction in the paper was based on the first order generalized Green-Schwarz transformation (\ref{gGS}) introduced in  \cite{MarquesNunez}. In order to proceed to even higher orders, we need further corrections to the generalized Green-Schwarz transformation. Interestingly, for the heterotic string these corrections are known non-perturbatively (through the so-called generalized Bergshoeff-de Roo identification), and perturbatively to second order in $\alpha'$ \cite{GenBdR}. Soon, an all-order proposal to corrections in the general bi-parametric case will appear \cite{BiParam}, where the  second-order corrections will be worked out explicitly. The strategy applied here, together with these results will permit to extend our computations to second order in $\alpha'$.

    \item {\bf Exceptional Drinfeld Doubles and maximal supergravity.} The results in this paper are at most compatible with half-maximal supergravity. Generalized dualities in the context of maximal supergravities gained renewed interest after the proposal for non-Abelian dualities of RR fields \cite{NARR}. Type II and M-theory give rise to rigid U-duality transformations upon compactifications on tori. Interestingly, the idea of generalized U-dualities was recently introduced in \cite{genU} and further discussed in \cite{othergenU}. Looking for higher order corrections to generalized U-dualities is out of reach at the moment, because these corrections are not even known in the Abelian case. There are promising steps in this direction \cite{Garousi}, systematics in the writing and counting of interactions is crucial \cite{EHS} because higher derivatives appear in maximal supergravity at order $\alpha'{}^3$, and so even the simplest corrections are hard to handle. Still, there is at the moment no higher derivative formulation of Exceptional Field Theory \cite{EFT} nor Type II DFT \cite{TypeII} (for a review see \cite{revEFT}), but generalized Scherk-Schwarz reductions have been extensively investigated \cite{gSSEFT} and surely constitute the proper framework to deal with generalized U-duality, in the same sense that Gauged DFT is the proper framework to deal with generalized T-duality.

\end{itemize}

We hope to make progress in these and other directions in the future.

{\bf Acknowledgements:} We thank R. Borsato, A. Catal-Ozer, F. Hassler, Y. Sakatani and L. Wulff for correspondence and discussions. Our work is supported by CONICET.

{\bf Note:} Upon completion of this work we became aware of \cite{Hassler:2020tvz} and \cite{Borsato:2020wwk} which overlap significantly with our results.

\end{document}